\documentclass{article}
\usepackage[utf8]{inputenc}
\usepackage[margin=1in]{geometry}
\usepackage{subcaption}
\usepackage{graphicx}
\usepackage{amsmath}
\usepackage[hyphens]{url}
\usepackage{hyperref}
\hypersetup{colorlinks=false,breaklinks=true,hidelinks}
\usepackage{cleveref}
\usepackage{apacite}
\usepackage{natbib}
\usepackage{authblk}
\usepackage{multirow}
\usepackage{booktabs}
\usepackage{makecell}
\graphicspath{{figures/}}
\usepackage{setspace}
\usepackage{pdfpages}

\title{California Earthquake Dataset for Machine Learning and Cloud Computing}
\author[1]{Weiqiang Zhu}
\author[1]{Haoyu Wang}
\author[1]{Bo Rong}
\author[2]{Ellen Yu}
\author[1]{Stephane Zuzlewski}
\author[2]{Gabrielle Tepp}
\author[1]{Taka'aki Taira}
\author[1]{Julien Marty}
\author[2]{Allen Husker}
\author[1]{Richard M Allen}
\affil[1]{Berkeley Seismological Laboratory, University of California, Berkeley, Berkeley, CA, USA}
\affil[2]{Caltech Seismological Laboratory, California Institute of Technology, Pasadena, CA, USA}

\date{}

\begin{document}

\maketitle

\begin{abstract}
The San Andreas Fault system, known for its frequent seismic activity, provides an extensive dataset for earthquake studies. The region's well-instrumented seismic networks have been crucial in advancing research on earthquake statistics, physics, and subsurface Earth structures.
In recent years, earthquake data from California has become increasingly valuable for deep learning applications, such as Generalized Phase Detection (GPD) for phase detection and polarity determination, and PhaseNet for phase arrival-time picking. The continuous accumulation of data, particularly those manually labeled by human analysts, serves as an essential resource for advancing both regional and global deep learning models.
To support the continued development of machine learning and data mining studies, we have compiled a unified California Earthquake Event Dataset (CEED) that integrates seismic records from the Northern California Earthquake Data Center (NCEDC) and the Southern California Earthquake Data Center (SCEDC). The dataset includes both automatically and manually determined parameters such as earthquake origin time, source location, P/S phase arrivals, first-motion polarities, and ground motion intensity measurements. The dataset is organized in an event-based format organized by year spanning from 2000 to 2024, facilitating cross-referencing with event catalogs and enabling continuous updates in future years.
This comprehensive open-access dataset is designed to support diverse applications including developing deep learning models, creating enhanced catalog products, and research into earthquake processes, fault zone structures, and seismic risks.
\end{abstract}

\section{Introduction}

Seismic catalogs play a pivotal role in advancing our understanding of earthquake processes and improving earthquake hazard assessments, particularly within the dynamic and complex San Andreas Fault System (SAFS).
By systematically recording and analyzing seismic events, earthquake catalogs provide a comprehensive framework for characterizing earthquake behavior, which is crucial for developing data-driven models of evolving seismicity and ground motion. The development of advanced earthquake detection algorithms has significantly improved the capability to identify smaller earthquakes, leading to more complete seismic catalogs. These enhanced datasets facilitate detailed studies of earthquake sequences and aftershock patterns, enable a deeper understanding of fault system interactions, and improve analyses of dynamic triggering and stress transfer processes, thereby offering critical insights into seismic activity and subsurface structures \citep{hauksson2012waveform,yang2012computing,brodsky2019importance,ross2019searching,park2022deep}.
Deep learning represents the state-of-the-art algorithm in artificial inelegance and has been widely adopted for seismic data processing \citep{perol2018convolutionala,ross2018generalized,zhu2019phasenet,mousavi2020earthquakea,zhu2022endtoend,mousavi2022deeplearninga}. These neural network models have been proven effective for building enhanced earthquake catalogs, studying complex earthquake sequences \citep{park2020machinelearningbasedb,liu2020rapida,tan2021machinea,park2021deepa,su2021high,wilding2022magmatic} and improving routine monitoring \citep{huang2020crowdquake,yeck2020leveraging,zhang2022loc,retailleau2022wrapper,shi2022malmi,tepp2024strategies}.
Compared to the traditional STA/LTA method, deep learning demonstrates higher sensitivity to weak signals from small earthquakes and greater robustness to noise spikes, thus detecting more events with fewer false positives.
Compared to template matching, deep learning generalizes similarity-based search without requiring precise seismic templates and operates significantly faster. Neural network models automatically learn to extract common features of earthquake signals from large training datasets, thereby gaining generalization capability for earthquakes beyond the training samples.

A key factor in the success of deep learning for earthquake detection and phase picking is the availability of extensive phase arrival-time measurements manually labeled by human analysts over several decades.
Many datasets have been compiled for training deep learning models at both global and regional scales.
Global datasets offer the advantage of encompassing a broad spectrum of waveforms and enhancing model generalization.
For example, the STanford EArthquake dataset (STEAD) \citep{mousavi2019stanforda} comprises $\sim$1.2 million seismic waveforms from local distances ($\leq$ 350 km); the Curated Regional Earthquake Waveforms (CREW) dataset \citep{suarez2024curated} contains $\sim$1.6 million waveforms from regional distances of 2 to 20 degrees; and the U.S. Geological Survey National Earthquake Information Center (NEIC) dataset includes $\sim$1.3 million seismic waveforms from global earthquakes spanning a wide range of magnitudes and distances \citep{yeck2020leveraginga}. Several other global benchmark datasets have been developed for deep learning models \citep{woollam2019convolutional,woollam2022seisbench}.
Additionally, multiple regional datasets have been compiled focusing on seismically active areas, such as the INSTANCE dataset of Italy \citep{michelini2021instance}, the DiTing dataset of China \citep{zhao2022diting}, the TXED dataset of Texas US \citep{chen2024txed}, and the Pacific Northwest dataset \citep{ni2023curated}.
Although two datasets have been developed for Northern California \citep{zhu2019phasenet} and Southern California \citep{ross2018generalized}, their formats differ significantly; for example, the Northern California dataset uses a window size of 30 seconds, whereas the Southern California dataset uses 3 seconds. Neither dataset has been updated with recent earthquake data.
There remains a need for a unified California earthquake dataset that is essential for advancing deep learning models for the entire San Andreas fault system.

In this work, we have combined earthquake catalogs and seismic waveforms from both Northern California (NC) and Southern California (SC) seismic networks to compile a comprehensive dataset that facilitates the continuous advancement of deep learning and cloud computing in seismology.
The dataset takes advantage of California's earthquake monitoring infrastructure, including its long monitoring duration and extensive network coverage, to provide robust and diverse data over time. The dataset also benefits from the high-quality earthquake catalogs of California's active seismic activity, which are systematically reviewed by analysts, to provide abundant well-labeled examples.
Its effectiveness in deep learning applications has been demonstrated through state-of-the-art models such as GPD \citep{ross2018generalized}, PhaseNet \citep{zhu2019phasenet}, and PhaseNO \citep{sun2023phase}.
The open accessibility of California seismic datasets on cloud platforms enhances reproducibility and broad usability worldwide.
This dataset is specifically designed to support training machine learning models, enabling robust model performance and facilitating large-scale seismic data mining. To ensure its ongoing relevance and reliability, the dataset will be regularly updated by incorporating data from future years, adding newly reviewed events to existing years, and improving data quality through community contributions and feedback. For the most up-to-date version of this dataset and accompanying paper, please refer to the arXiv version.%

\section{Event Dataset}

\subsection{Earthquake catalogs}

\begin{figure}
    \centering
    \begin{subfigure}{0.42\textwidth}
    \includegraphics[width=\textwidth]{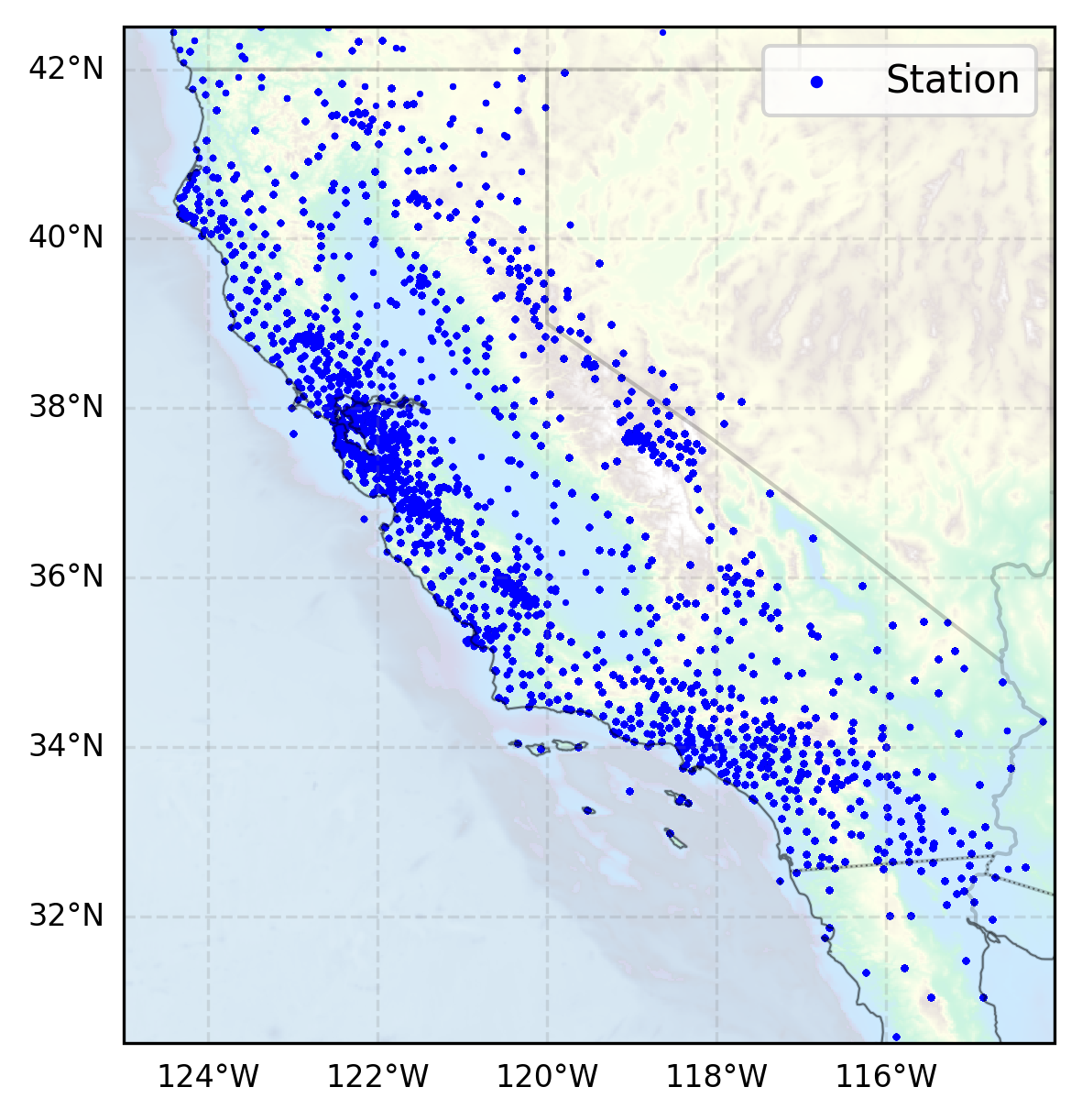}
    \caption{}
    \end{subfigure}
    \begin{subfigure}{0.5\textwidth}
    \includegraphics[width=\textwidth]{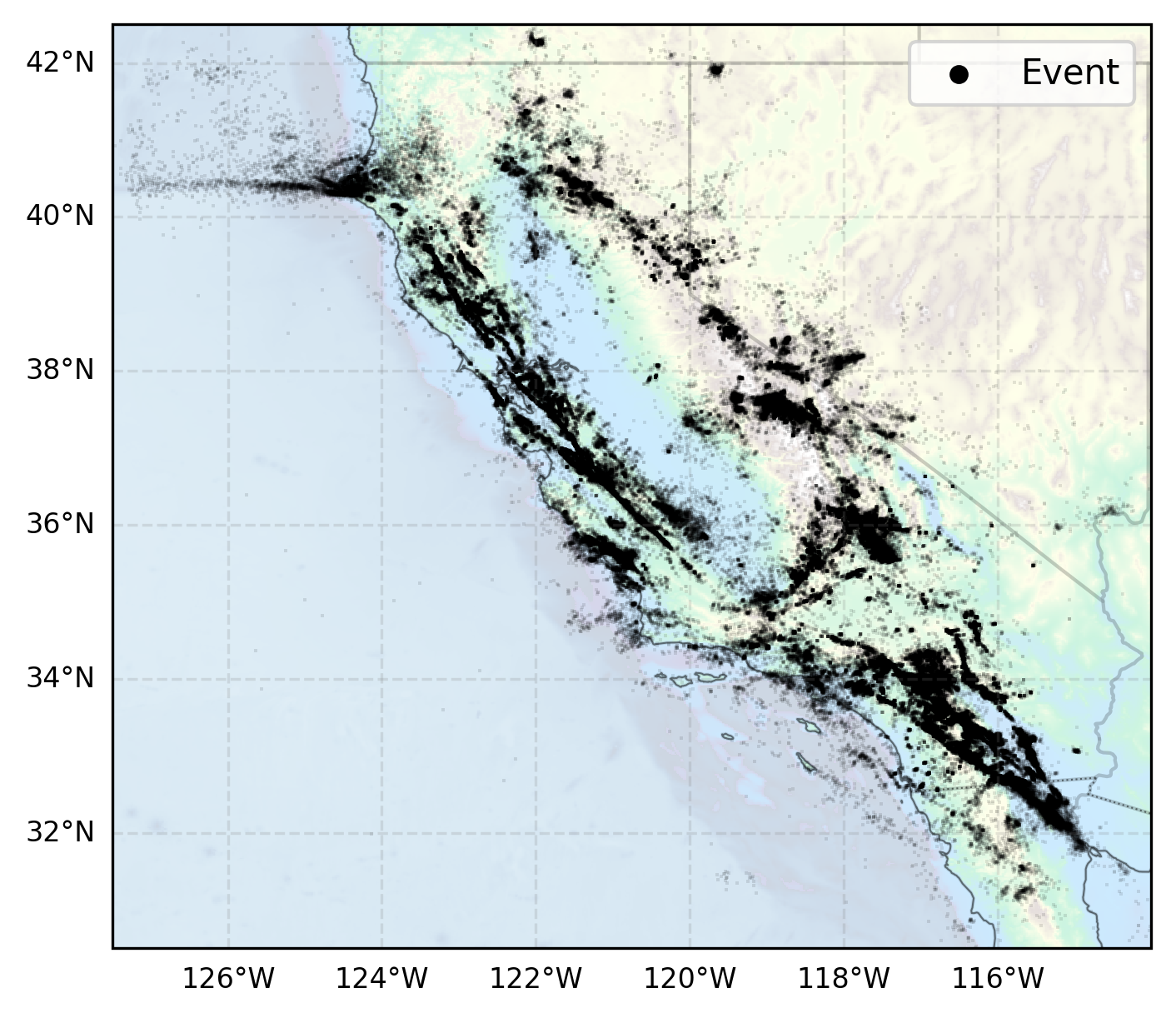}
    \caption{}
    \end{subfigure}
    \caption{Spatial distribution of selected (a) seismic stations and (b) earthquakes in California with manual labels in the CEED dataset.}
    \label{fig:event_station}
    \end{figure}

\begin{figure}
    \centering
    \begin{subfigure}{0.48\textwidth}
    \includegraphics[width=\textwidth]{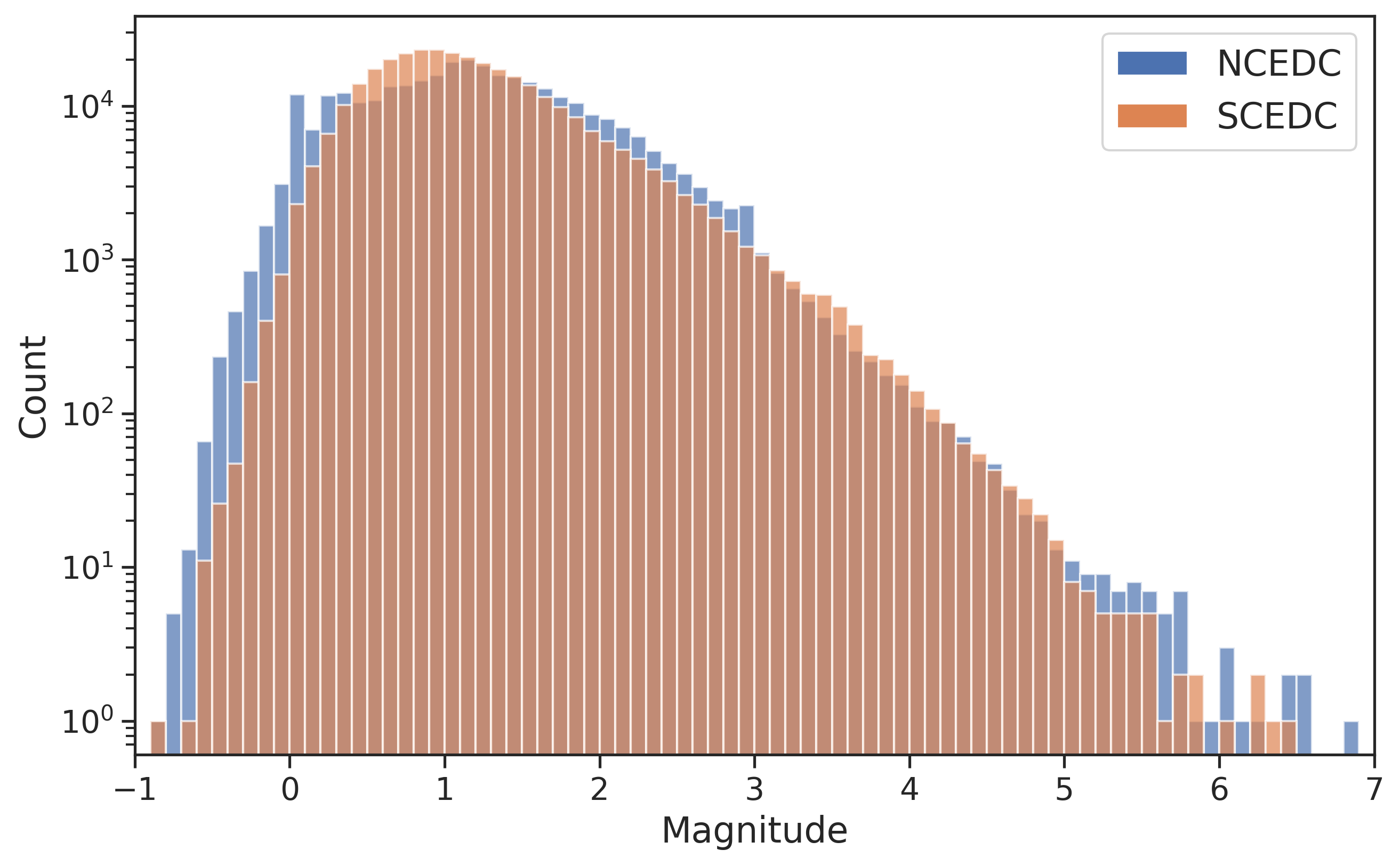}
    \caption{}
    \end{subfigure}
    \begin{subfigure}{0.48\textwidth}
    \includegraphics[width=\textwidth]{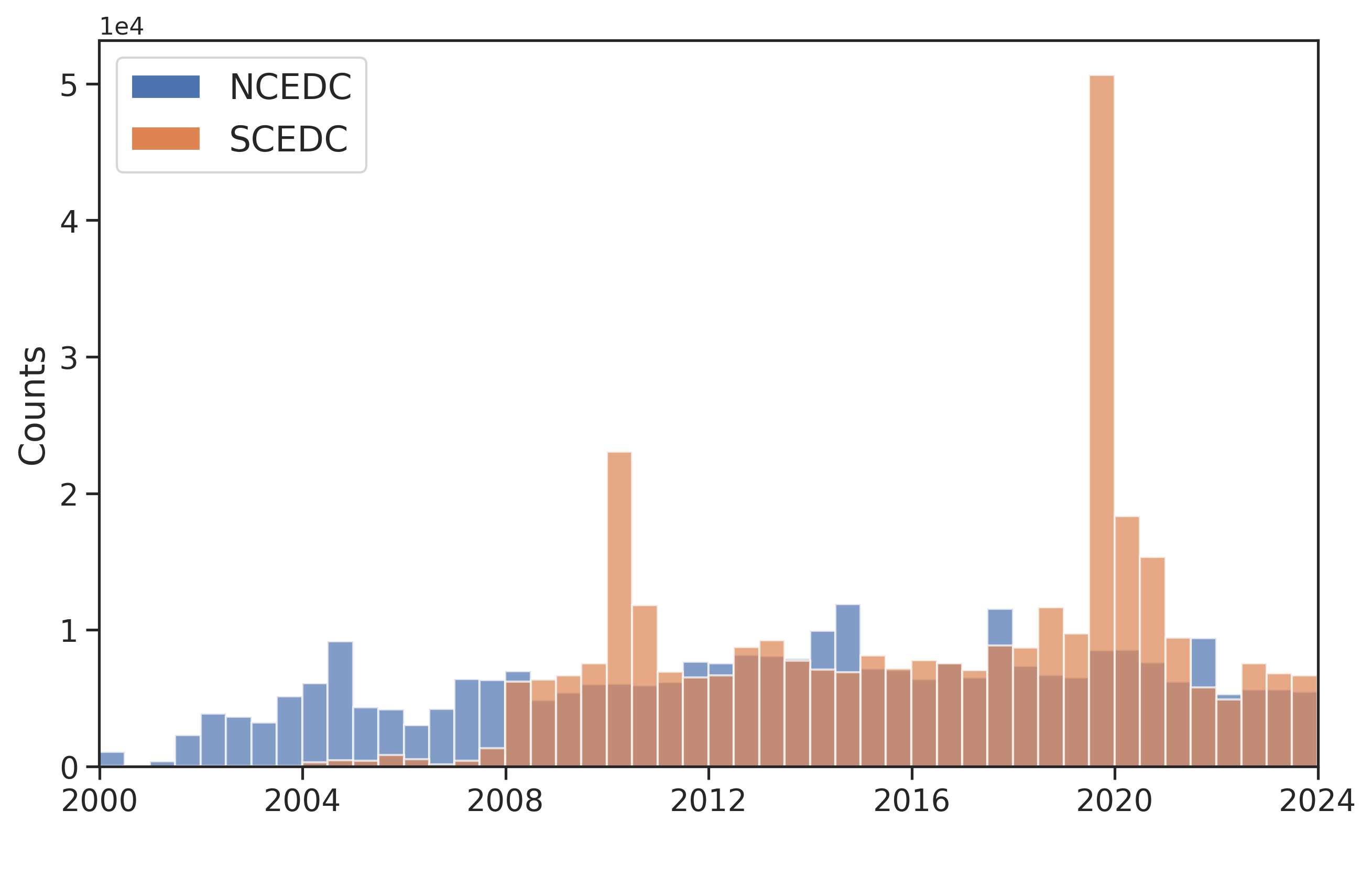}
    \caption{}
    \end{subfigure}
    \caption{Distribution of (a) earthquake magnitudes and (b) origin times.}
    \label{fig:mag_time}
    \end{figure}

Based on earthquake catalogs and continuous waveforms from Northern and Southern California as of 2023 (\Cref{fig:event_station}), we have compiled 325K events and 1.1M three-component waveform samples with manually reviewed labels from the Northern California Earthquake Data Center (NCEDC) and 328K events and 3.0M waveform samples from the Southern California Earthquake Data Center (SCEDC), respectively \citep{ncedc2014northern,scedc2013southern}. We retain only waveform containing both P and S picks in the dataset, which explains the larger number of waveform samples from SCEDC compared to NCEDC.
The distributions of events and station coverage are shown in \Cref{fig:event_station}. The selected stations provide comprehensive coverage across California, capturing diverse seismic activities. These include predominantly tectonic earthquakes, induced seismicity in geothermal fields (e.g., The Geysers, Coso, and Salton Sea), volcanic earthquakes (e.g., Long Valley, Lassen Peak, and Mount Shasta), and offshore events at the Mendocino Triple Junction.
The events span a magnitude range from M1 to M7 over approximately two decades (\Cref{fig:mag_time}).
This extensive variety of seismic events helps enhance the generalization capability of deep learning models.


\subsection{Pre-processing}

We applied minimal pre-processing to the dataset, including removing the mean, resampling to 100 Hz, rotating to ENZ directions, and converting to physical units of velocity or acceleration.
We chose not to remove station response because reversing this process can be challenging. Instead, we preserved the station response files to enable straightforward removal of instrument response as needed for specific tasks.
We maintained the inherent complexity of waveforms in the training dataset to enable models to learn processing under various challenging conditions. Machine learning models need to be robust in analyzing continuous seismic archives that often present challenging and unexpected data issues, including missing data, data drifting, abrupt changes, instrument noise, and anthropogenic noise.
A remaining challenge is the presence of incorrect labels in standard catalogs, which can adversely affect model training and performance. Such labeling errors are inevitable given the extensive scale and duration of the California earthquake catalogs.
A related issue is the absence of labels for non-cataloged events within the waveform window.
We have not applied filtering to identify and exclude these incorrect labels, reserving this for future work.
One potential solution is to apply trained deep learning models to identify potentially incorrect manual labels and detect missing events in the dataset, followed by manual verification and correction to enable continuous improvement of dataset quality alongside model development.
We will also rely on community efforts to enhance dataset quality. Data issues and new labels can be reported directly on the dataset repository at \url{https://huggingface.co/datasets/AI4EPS/CEED/discussions}

\subsection{Dataset statistics}

\begin{figure}
    \centering
    \includegraphics[width=0.8\textwidth]{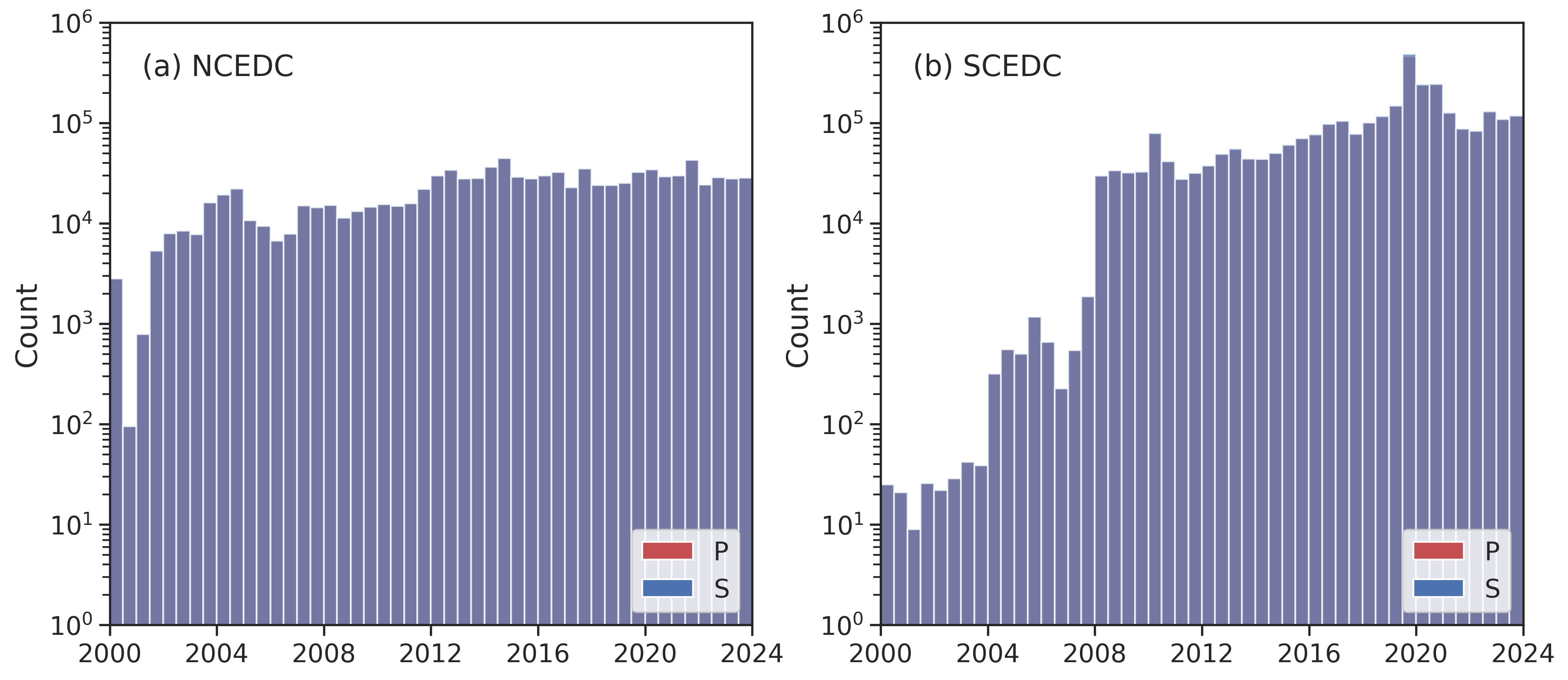}
    \caption{Temporal distribution of pairs of P and S phase arrival picks in the CEED dataset, which contains 1.1M and 3.0M pairs of P and S picks from NCEDC and SCEDC, respectively, as of 2023.}
    \label{fig:phase_arrival}
\end{figure}

\begin{figure}
    \centering
    \includegraphics[width=0.6\textwidth]{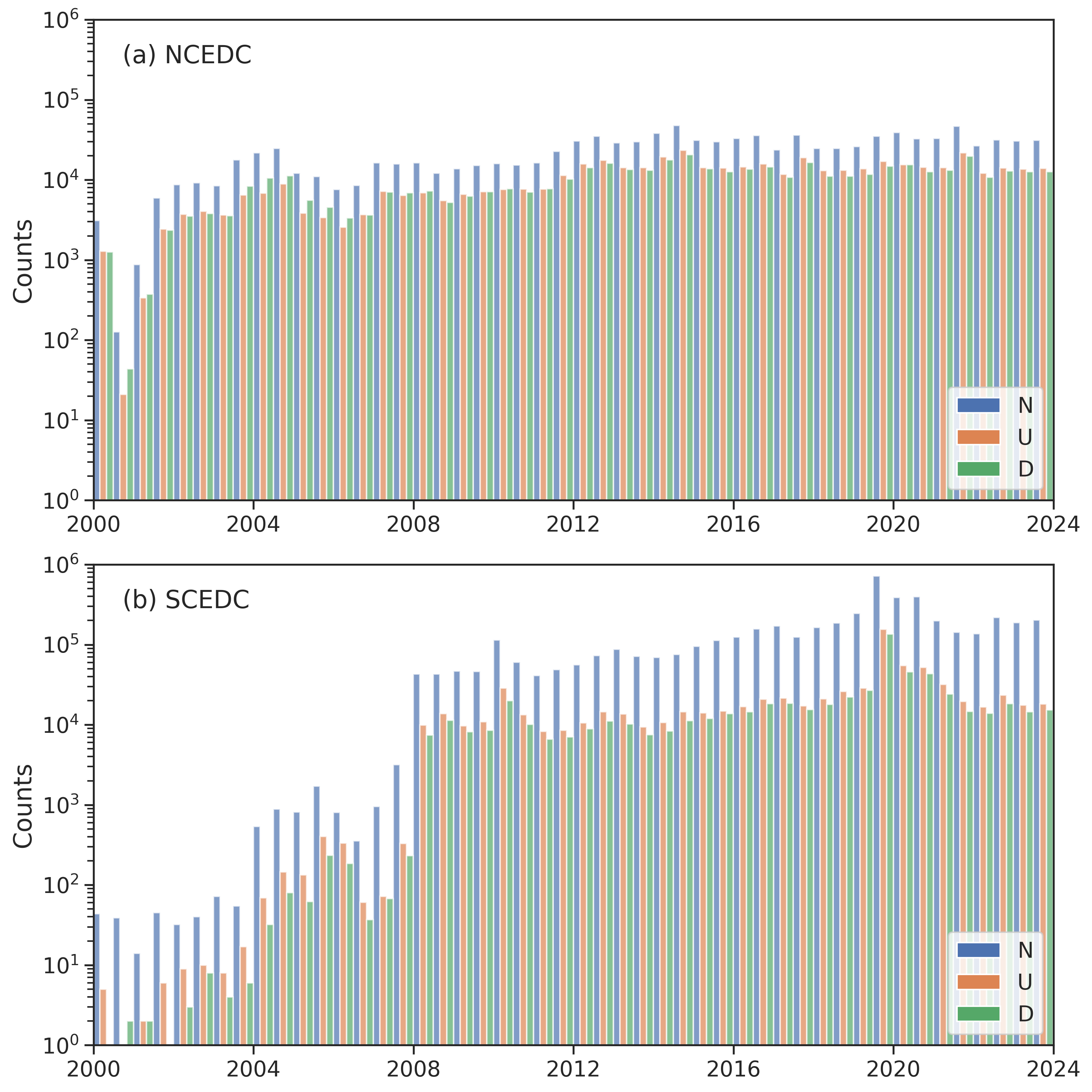}
    \caption{Distribution of first-motion polarity picks in the CEED dataset. Polarities are classified as up (``U''), down (``D''), or unknown (``N''). The dataset includes 1.0M million and 1.4M definitive (``U'' and ``D'') polarity picks from NCEDC and SCEDC, respectively, as of 2023.}
    \label{fig:phase_polarity}
\end{figure}

\begin{figure}
    \centering
    \begin{subfigure}{0.48\textwidth}
    \includegraphics[width=\textwidth]{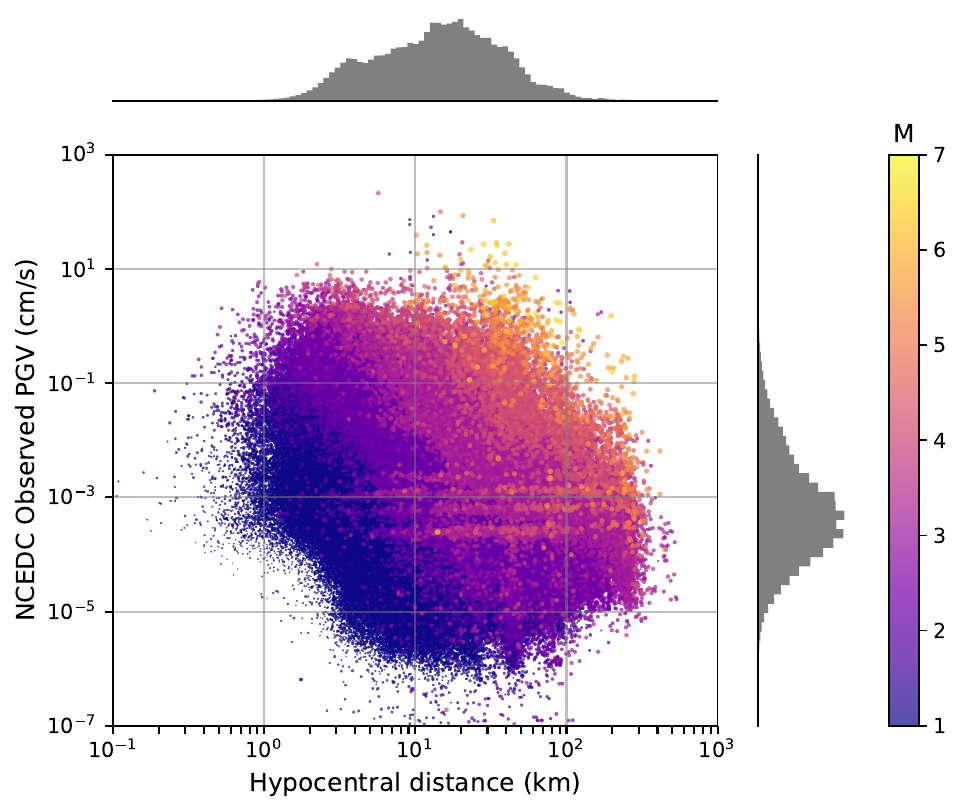}
    \caption{}
    \end{subfigure}
    \begin{subfigure}{0.48\textwidth}
    \includegraphics[width=\textwidth]{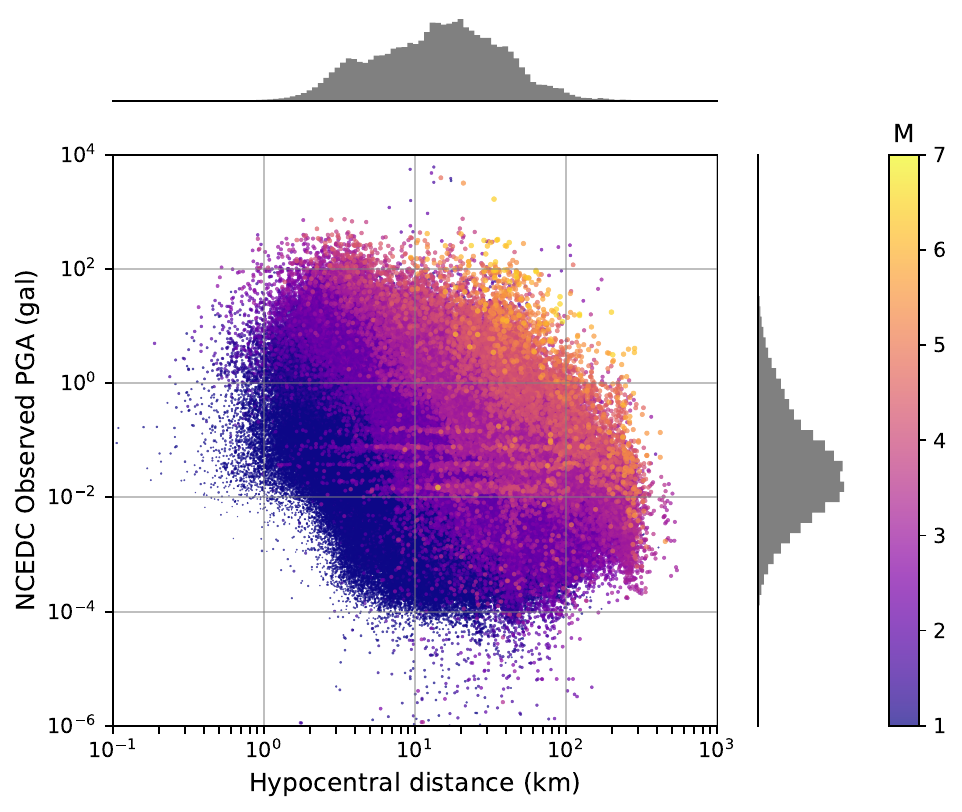}
    \caption{}
    \end{subfigure}
    \centering
    \begin{subfigure}{0.48\textwidth}
        \includegraphics[width=\textwidth]{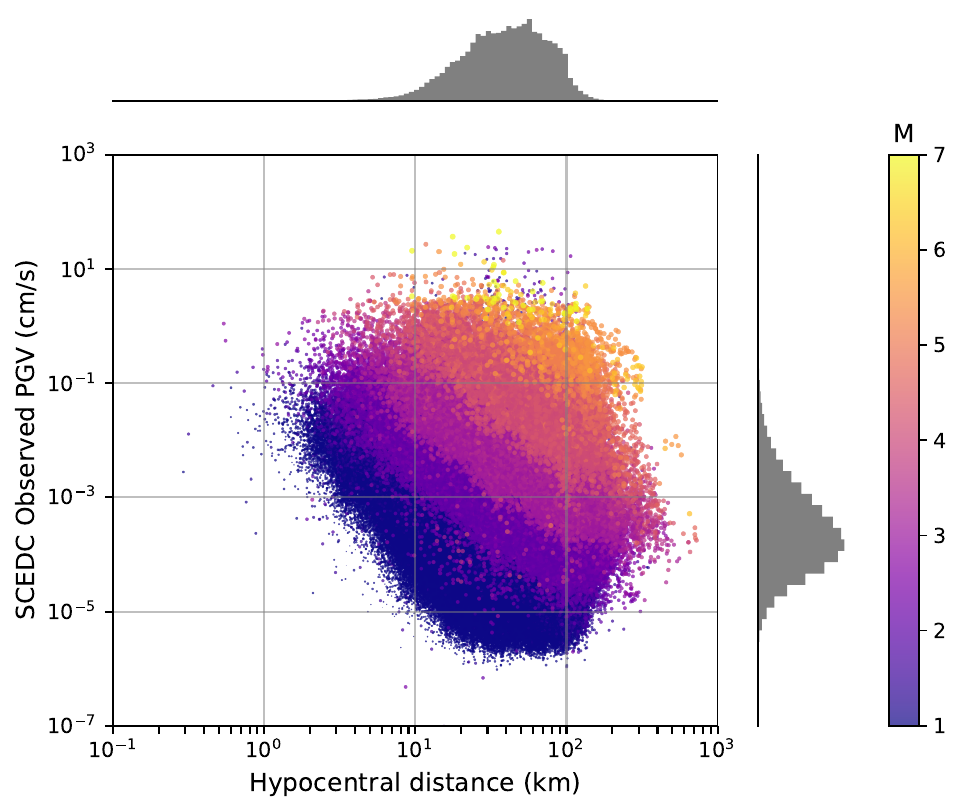}
        \caption{}
    \end{subfigure}
    \begin{subfigure}{0.48\textwidth}
        \includegraphics[width=\textwidth]{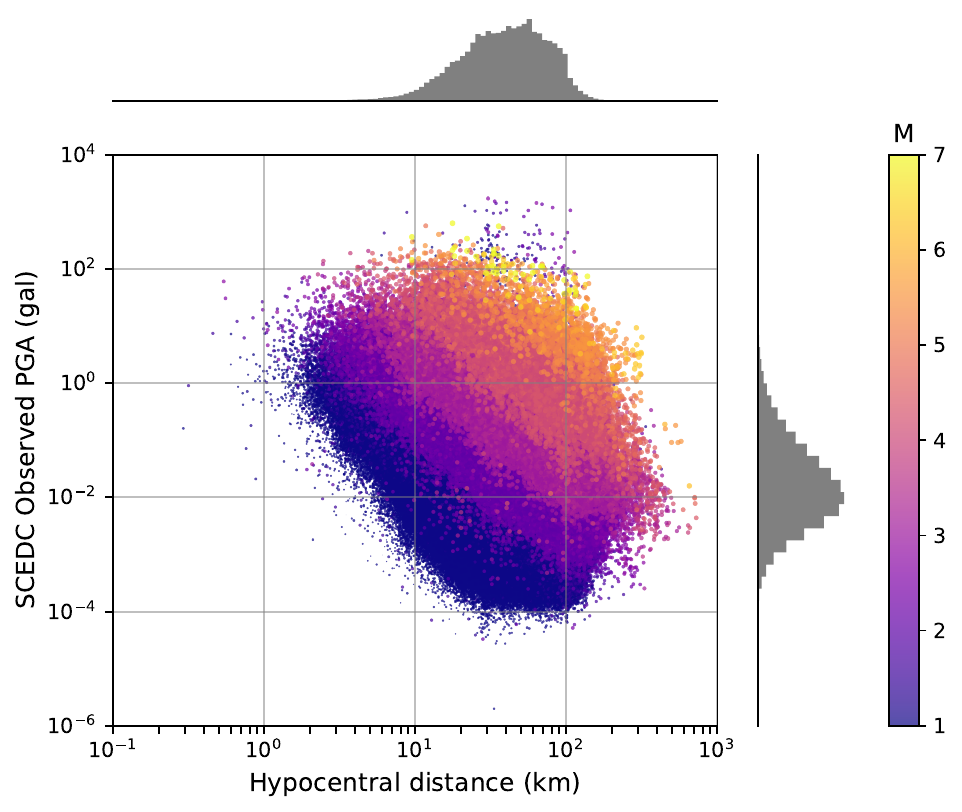}
        \caption{}
    \end{subfigure}
    \caption{Distribution of ground motion intensity measurements in the CEED dataset: (a) NCEDC Peak Ground Velocity (PGV), (b) NCEDC Peak Ground Acceleration (PGA), (c) SCEDC Peak Ground Velocity (PGV), and (d) SCEDC Peak Ground Acceleration (PGA). The dataset includes 1.1M and 3.1M measurements from NCEDC and SCEDC, respectively, through 2023}
    \label{fig:ground_motion}
\end{figure}

\begin{figure}
    \centering
    \begin{subfigure}{0.43\textwidth}
    \includegraphics[width=\textwidth]{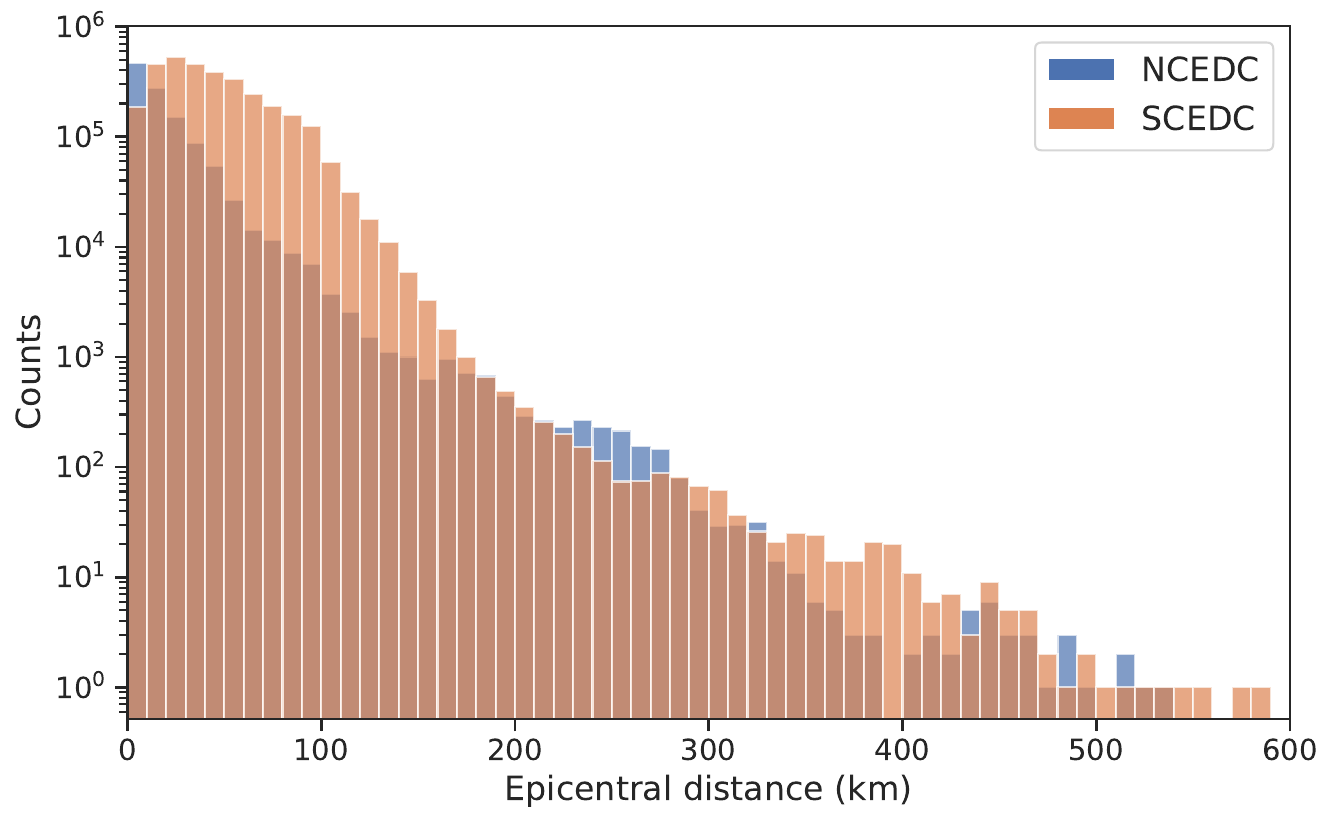}
    \caption{}
    \end{subfigure}
    \begin{subfigure}{0.43\textwidth}
    \includegraphics[width=\textwidth]{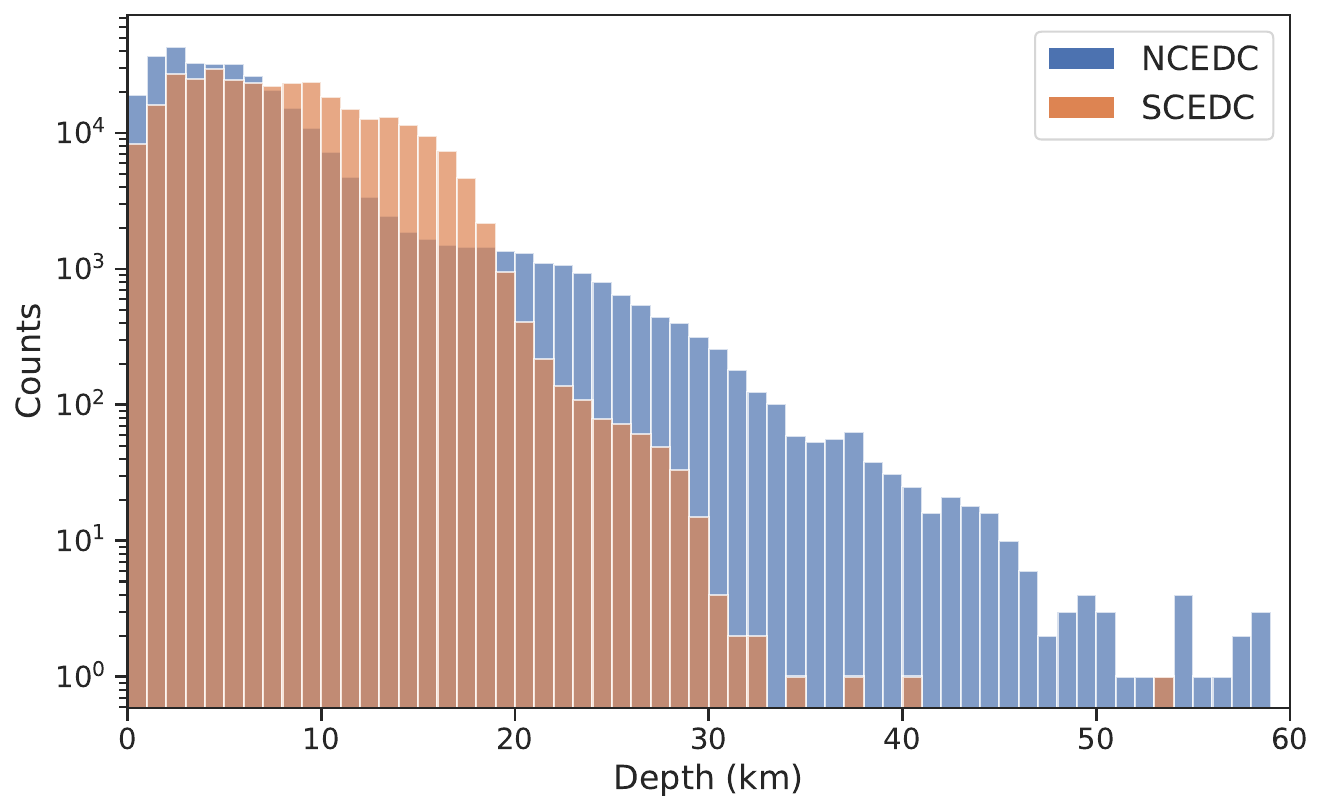}
    \caption{}
    \end{subfigure}
    \begin{subfigure}{0.43\textwidth}
    \includegraphics[width=\textwidth]{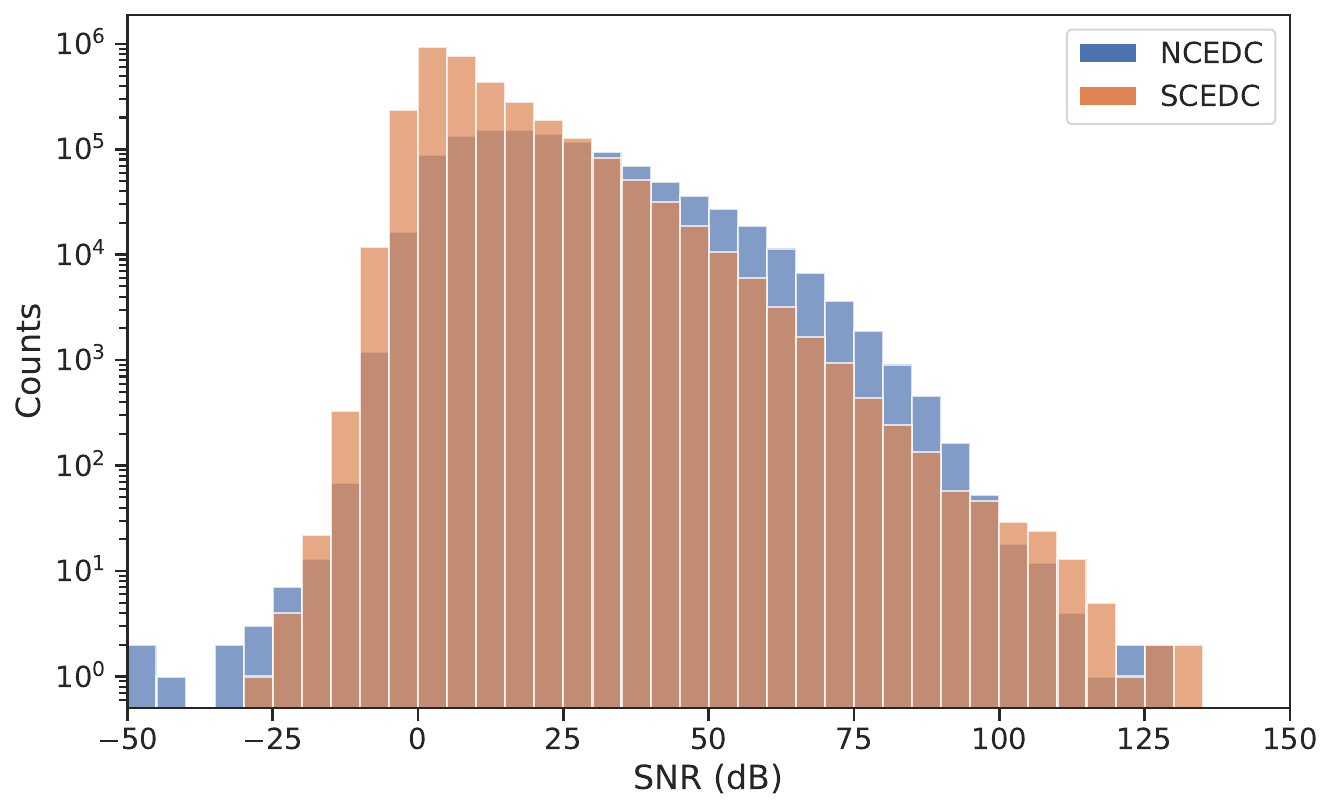}
    \caption{}
    \end{subfigure}
    \begin{subfigure}{0.43\textwidth}
    \includegraphics[width=\textwidth]{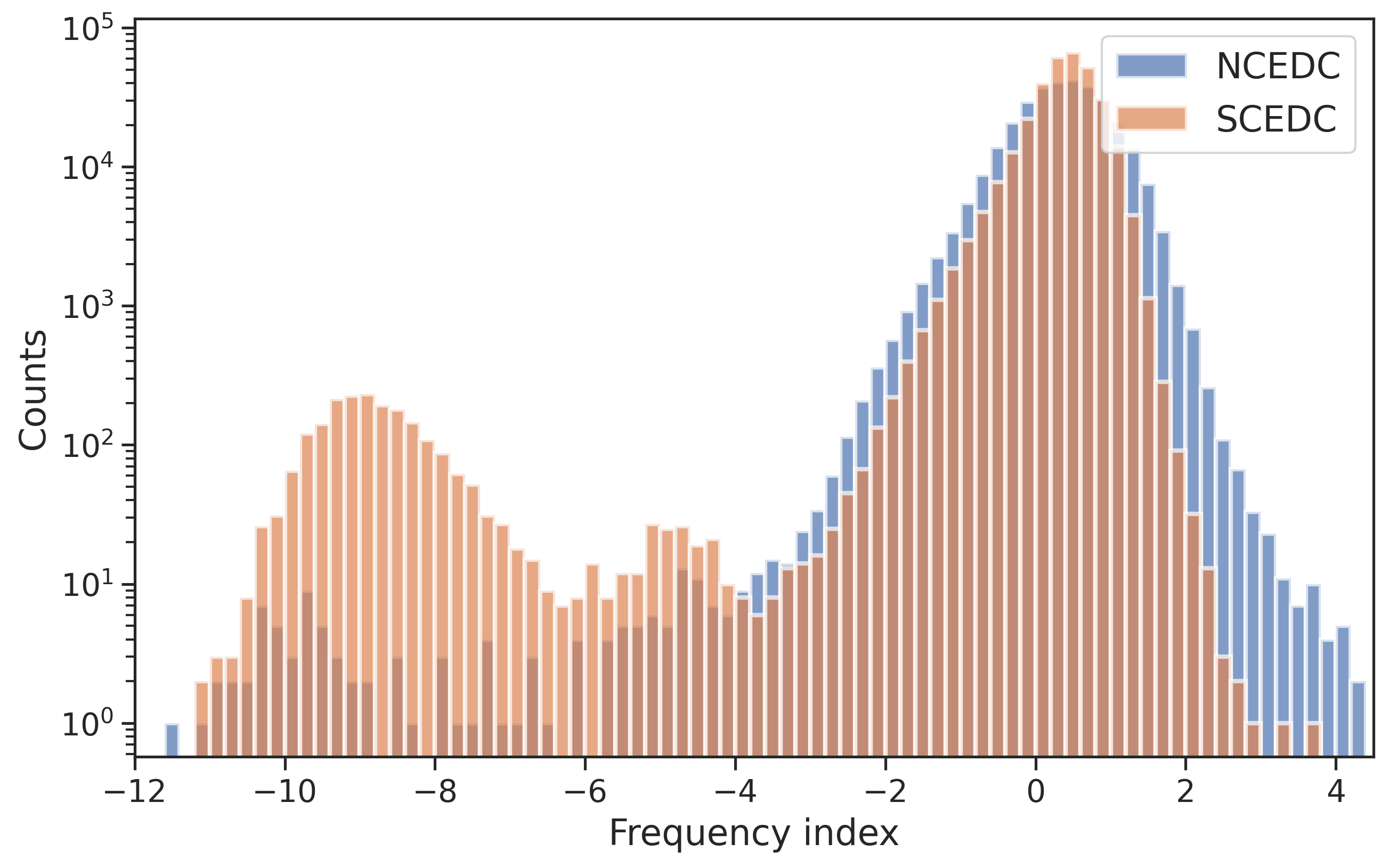}
    \caption{}
    \end{subfigure}
    \begin{subfigure}{0.6\textwidth}
    \includegraphics[width=\textwidth]{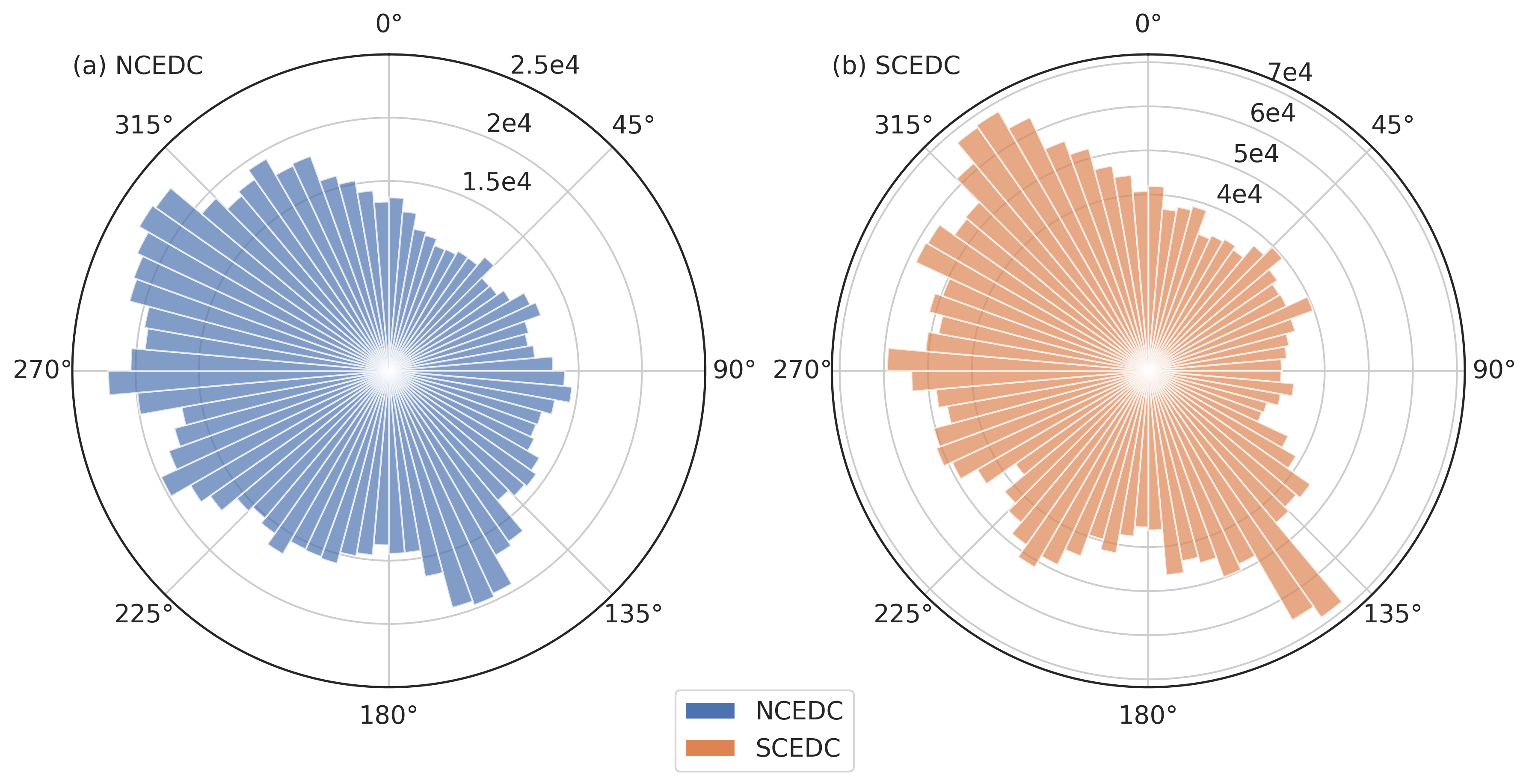}
    \caption{}
    \end{subfigure}
    \caption{Key characteristics of seismic events and waveforms in the CEED dataset: (a) distribution of epicentral distances, (b) distribution of event depths, (c) distribution of signal-to-noise ratios (SNR), (d) distribution of frequency indices, and (e) distribution of back azimuths. These distributions highlight the dataset's coverage of diverse seismic recording conditions.}
    \label{fig:dist_depth_snr}
\end{figure}

\begin{figure}
    \centering
    \begin{subfigure}{\textwidth}
    \includegraphics[width=\textwidth]{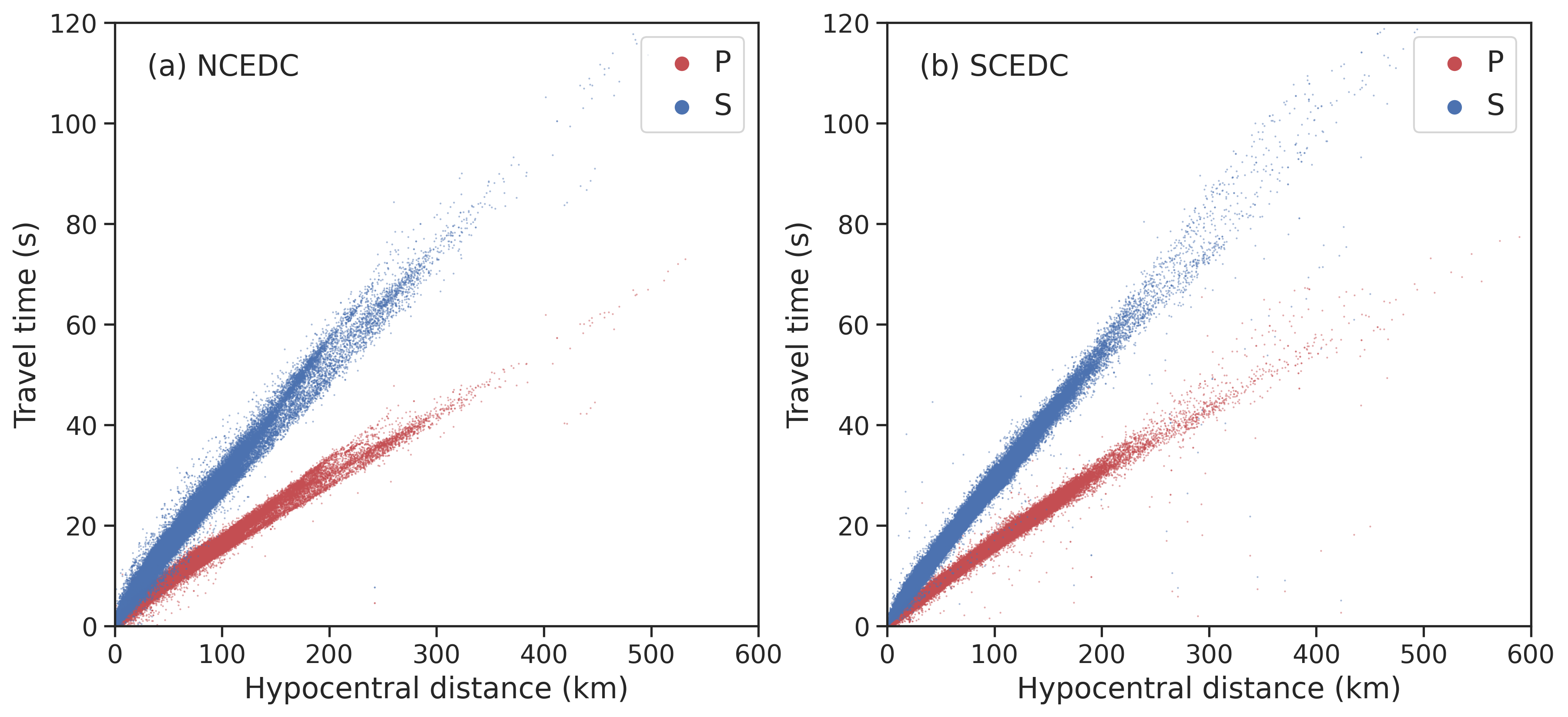}
    \caption{}
    \end{subfigure}
    \caption{Relationship between seismic wave travel times and hypocentral distances.}
    \label{fig:travel_time_distance}
    \end{figure}

\begin{figure}
    \centering
    \begin{subfigure}{0.8\textwidth}
    \includegraphics[width=\textwidth]{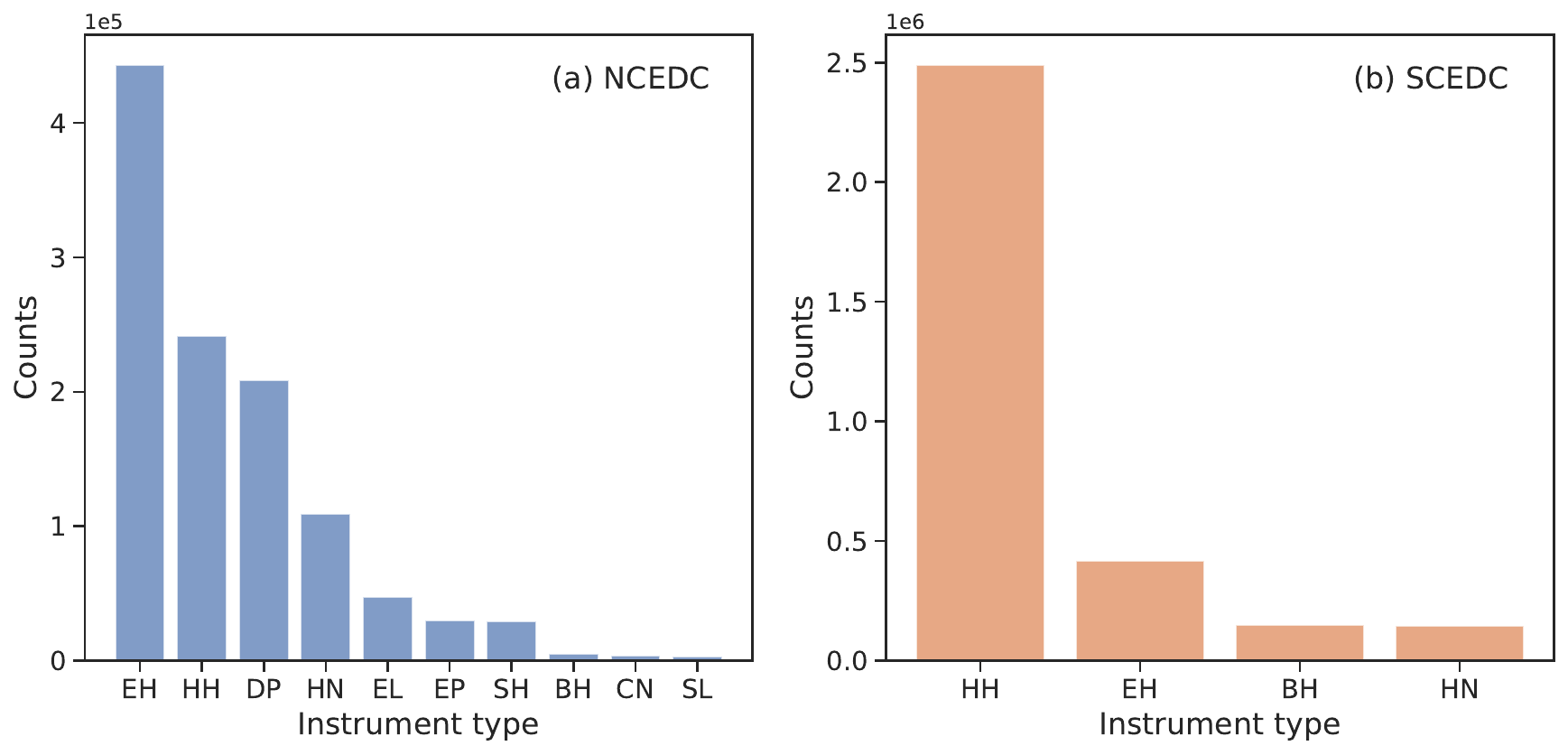}
    \caption{}
    \end{subfigure}
    \caption{Major seismometer instrument types in the CEED dataset. Definitions of these instrument type codes can be found in the documentation of the International Federation of Digital Seismograph Networks (FDSN) at \url{https://docs.fdsn.org/projects/source-identifiers/en/v1.0/channel-codes.html}.}
    \label{fig:instrument}
\end{figure}

To support diverse machine learning applications, we have included labels of phase arrival-times, first-motion polarities, and ground motions (PGA and PGV), and basic earthquake source information.
\Cref{fig:phase_arrival,fig:phase_polarity,fig:ground_motion} show the distribution of these labels for Northern and Southern California.
The dataset encompasses over 4.1M waveforms with labels, making it one of the largest datasets for machine learning through 2023.
\Cref{fig:dist_depth_snr,fig:travel_time_distance,fig:instrument} show the distributions of epicentral distance, source depths, signal-to-noise ratios (SNRs), frequency index, back azimuth, travel times, and instrument types.
The frequency index is computed based on the ratio of dominant frequency bands between 1–5 Hz and 10–15 Hz \citep{buurman2006seismic,zhong2024deep}.
Due to the inherently imbalanced nature of these distributions, careful consideration is necessary when using the dataset for training deep learning models, whose performance strongly depend on the distribution of training and test datasets.
While these specific data distributions may not significantly impact applications within California, they could limit model generalization to other global regions and different earthquake types.
Data augmentation techniques, such as oversampling or downsampling, could help improve model generalization \citep{zhu2020seismic}.

\subsection{Dataset format}

\begin{figure}
    \centering
    \begin{subfigure}{0.43\textwidth}
    \includegraphics[width=\textwidth]{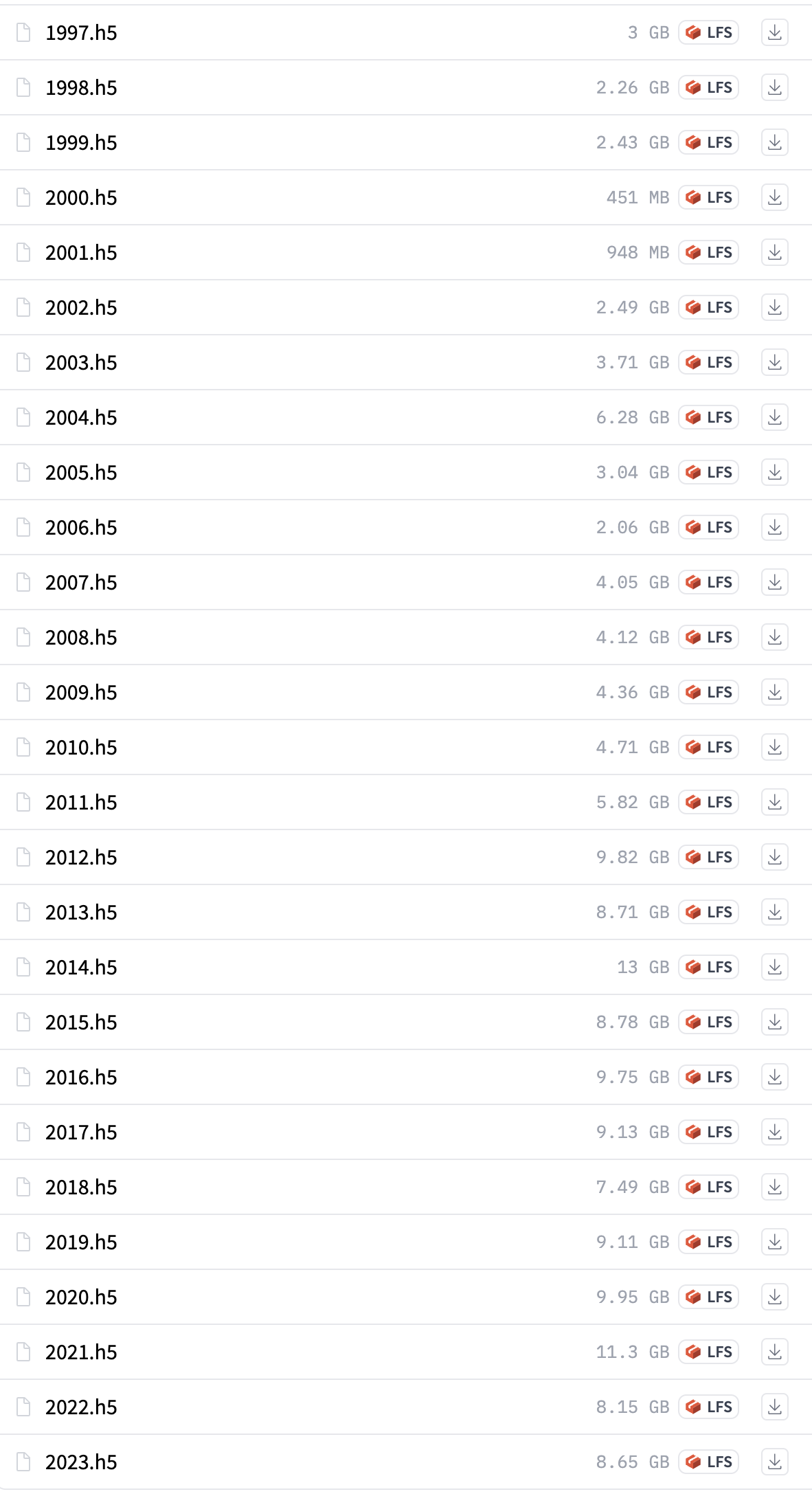}
    \caption{}
    \end{subfigure}
    \begin{subfigure}{0.43\textwidth}
    \includegraphics[width=\textwidth]{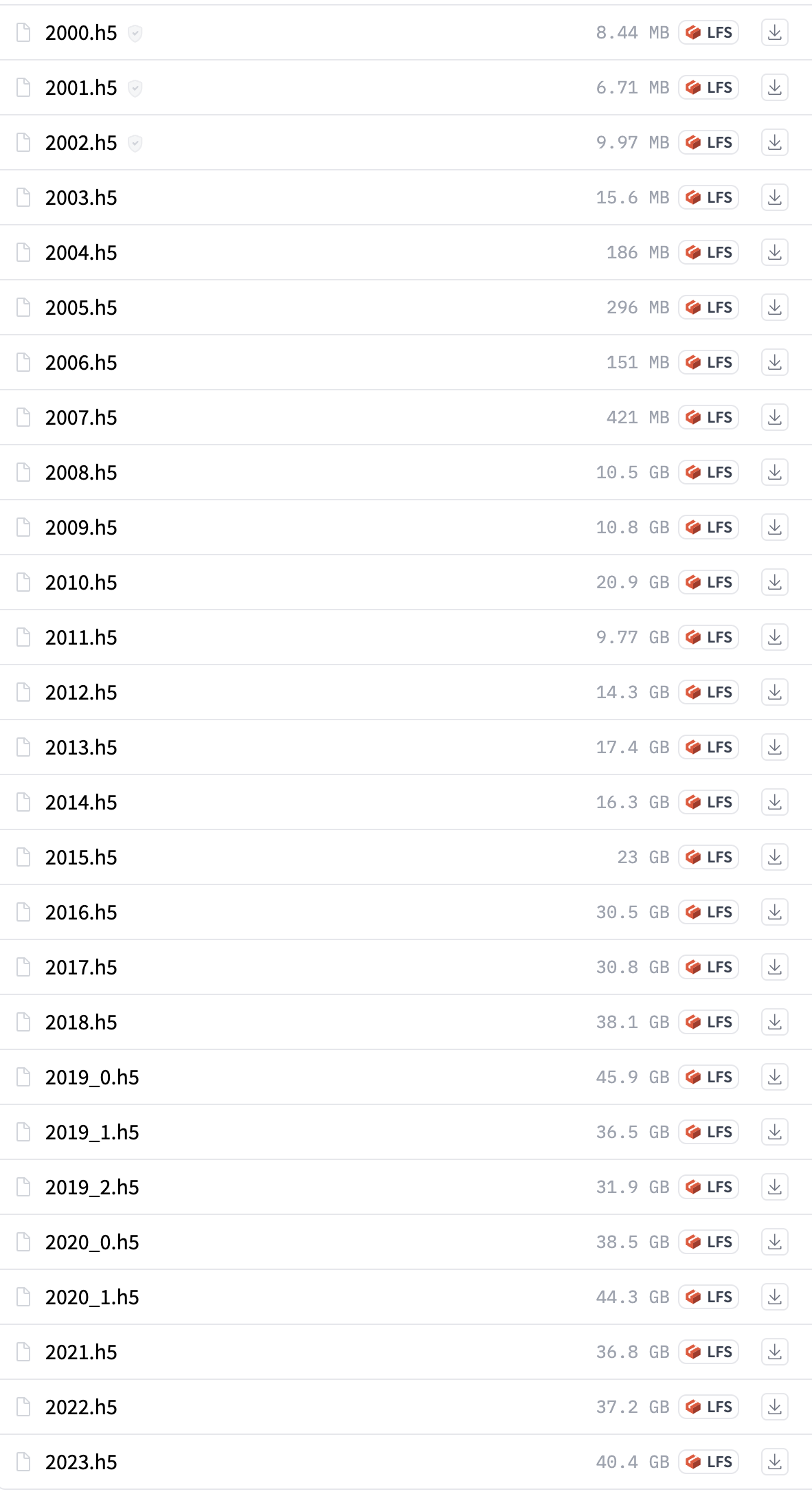}
    \caption{}
    \end{subfigure}
    \caption{Organization of yearly data available in the CEED dataset for (a) NCEDC and (b) SCEDC. Seismic event waveforms are organized chronologically in separate HDF5 files by year, enabling efficient data access and straightforward updates. The internal structure of each HDF5 file follows a standardized format detailed in \Cref{fig:h5_format}.}
    \label{fig:huggingface}
\end{figure}

\begin{figure}
    \centering
    \includegraphics[width=0.8\textwidth]{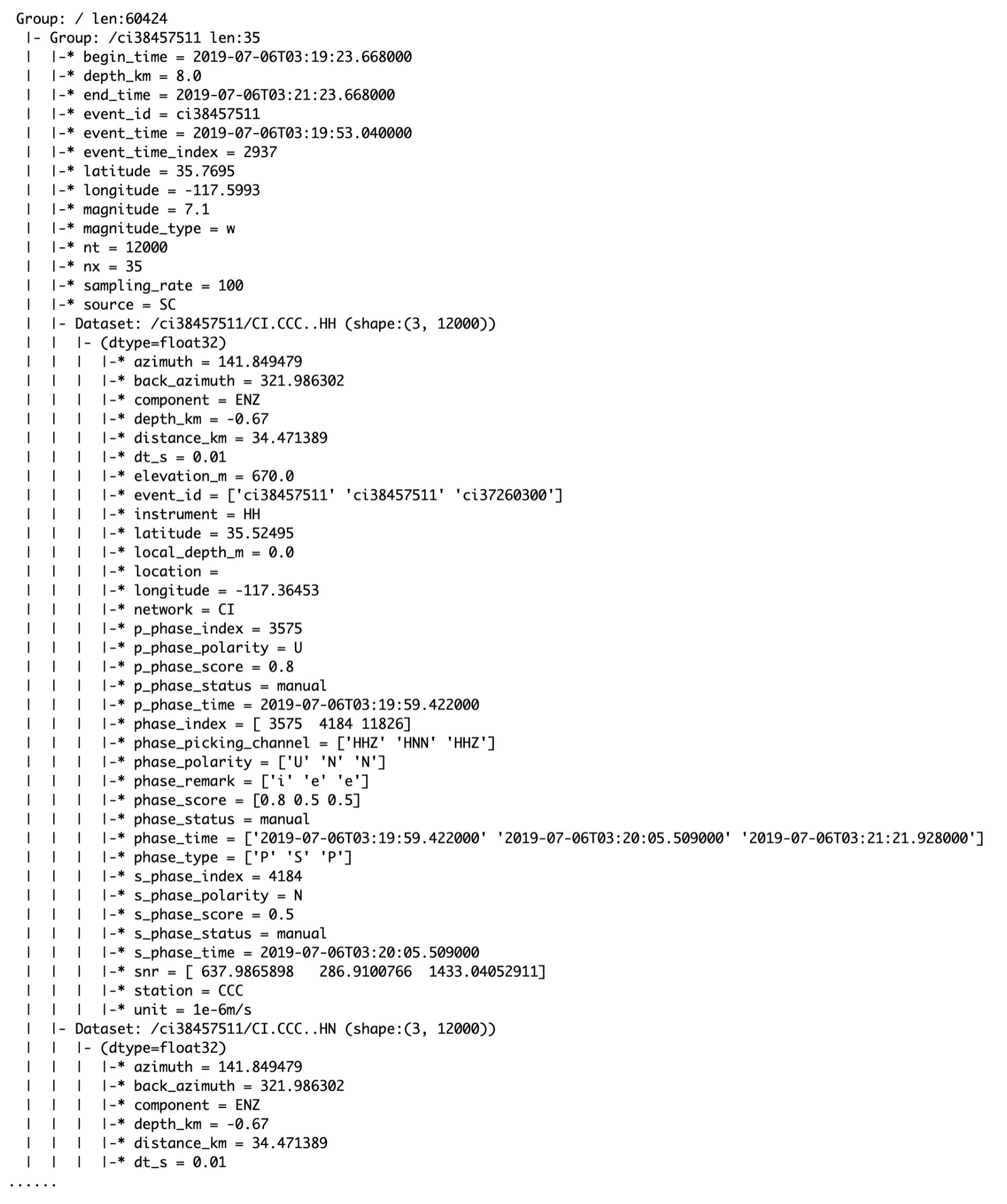}
    \caption{Structure of the hierarchical HDF5 format used in the CEED dataset. The event-based organization enables efficient data access and cross-referencing with the USGS ComCat system (example: \url{https://earthquake.usgs.gov/earthquakes/eventpage/ci38457511/origin/phase}). The format supports both single-station and network-based machine learning applications.}
    \label{fig:h5_format}
\end{figure}

Most existing datasets built for deep learning are organized by individual waveform samples \citep{ross2018generalized,zhu2019phasenet,zhao2022diting,mousavi2019stanforda,woollam2019convolutional,yeck2020leveraginga,woollam2022seisbench,ni2023curated}. This approach, while sufficient for training single-station-based deep learning models such as phase picking models based on three-component waveforms, is not optimal for training multi-station or network-based deep learning models, which have demonstrated improvements over single-station models \citep{sun2023phase,si2024all,feng2022edgephase}.
To ensure compatibility with both single-station-based and multi-station-based models, we have adopted a hierarchical event-based format for the dataset.
The dataset is organized by years to facilitate continuous updates (\Cref{fig:huggingface}). Within the HDF5 dataset of each year, waveforms are organized by event IDs and then by station IDs (\Cref{fig:h5_format}).
This format enables straightforward cross-referencing with the USGS Comprehensive Earthquake Catalog (ComCat) \citep{us2017advanced}, such as \url{https://earthquake.usgs.gov/earthquakes/eventpage/ci38457511}.
The HDF5 format includes comprehensive catalog and phase information. Event source information, such as origin time, location, magnitude, and mechanism, is stored in the attributes of the event group level (e.g., ``ci38457511"). For each event, the corresponding station and label information, including network code, station code, epicenteral distance, back azimuth, and labels of phase picks (arrivals, polarities, and PGA/PGV), is stored in the attributes of the waveform dataset level (e.g., "ci38457511/CI.CCC..HH").
This design not only simplifies data addition and updates but also provides flexibility to incorporate new attributes at both event and station levels for training both single-station-based and multi-station-based deep learning models.

\section{Applications}

\subsection{Machine learning}

The California earthquake event dataset serves as a primary resource for training deep learning models.
The dataset is hosted on Hugging Face, a leading platform for public datasets and models (\url{https://huggingface.co/datasets/AI4EPS/CEED}).
Users can easily access the dataset using \textit{git}\footnote{https://huggingface.co/datasets/AI4EPS/CEED?clone=true} or the \textit{datasets}\footnote{https://huggingface.co/docs/datasets} package provided by Hugging Face, as demonstrated in the notebook examples in the supplementary materials.
A portion of the dataset predating 2018 has already been successfully utilized in training GPD and PhaseNet.
The newly added data from subsequent years can further enhance these models and support the development of more advanced approaches.
For example, the PhaseNet+ model employs both phase arrival picks and polarity picks to train a multitask deep learning model for constraining both earthquake locations and focal mechanisms \citep{zhu2024end}; the QuakeFormer model utilizes PGV and PGA measurements to develop a non-ergodic ground motion prediction model for California \citep{feng2024uniform}; and the PhaseNO and EQNet models \citep{sun2023phase,zhu2024end} leverage the event-based format to develop multi-station-based phase picking models that enhance small earthquake detection sensitivity while suppressing false positive picks.
The open-access dataset could ensure reproducibility and foster collaboration, serving as a valuable resource for training and benchmarking machine learning models across diverse earthquake types throughout California. 
In conjunction with datasets collected in other seismically active regions, such as Alaska, Japan, and Italy, the California dataset can further contribute to the development of global deep learning models.

\subsection{Cloud computing}

\begin{figure}
    \centering
    \includegraphics[width=0.6\textwidth]{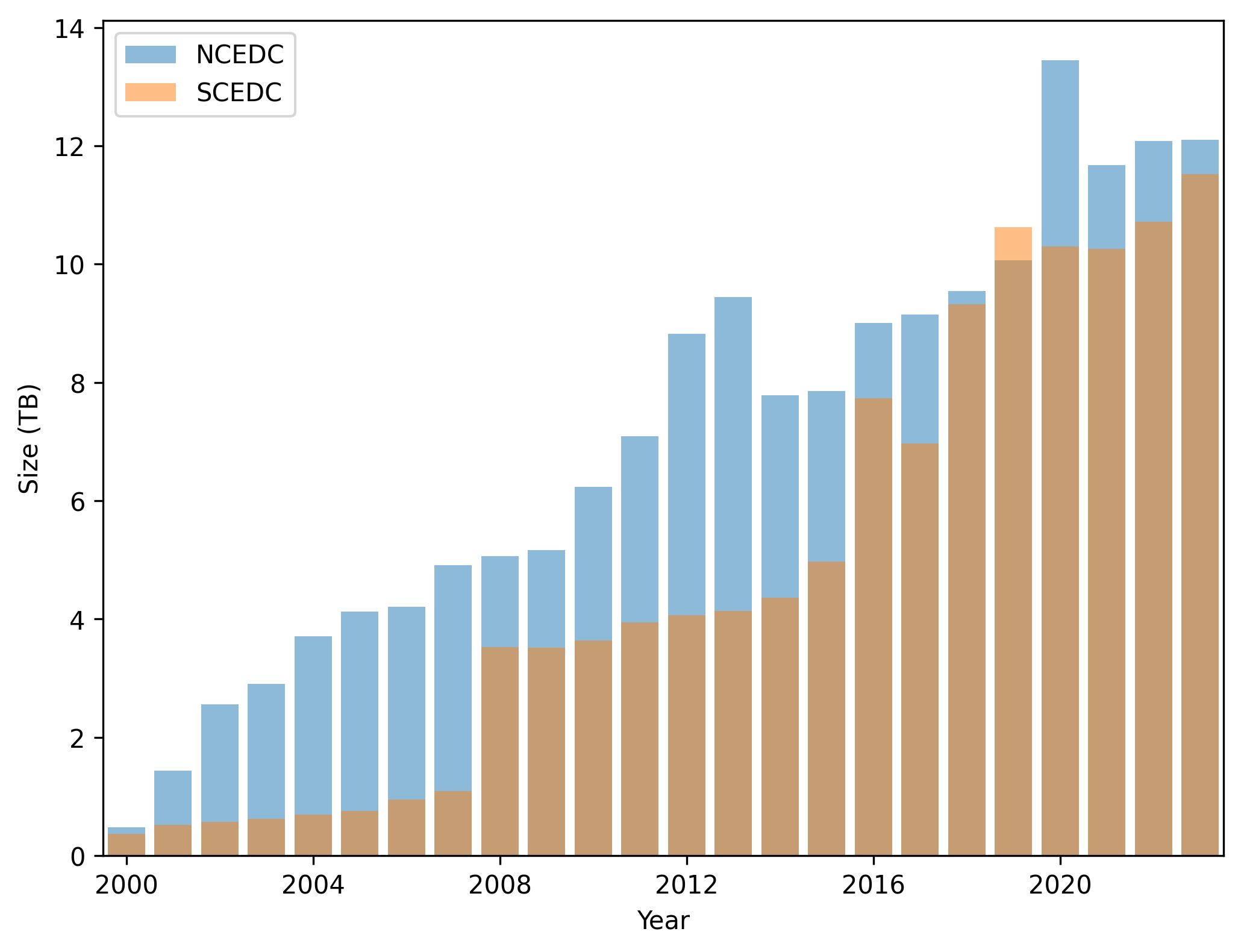}
    \caption{Cloud-hosted seismic data volume available for each year maintained by NCEDC and SCEDC. Detailed storage statistics and access information are available at \url{https://ncedc.org/db/cloud.html} and \url{https://scedc.caltech.edu/data/cloud.html}.}
    \label{fig:aws_bucket}
    \end{figure}

\begin{figure}
    \centering
    \includegraphics[width=0.6\textwidth]{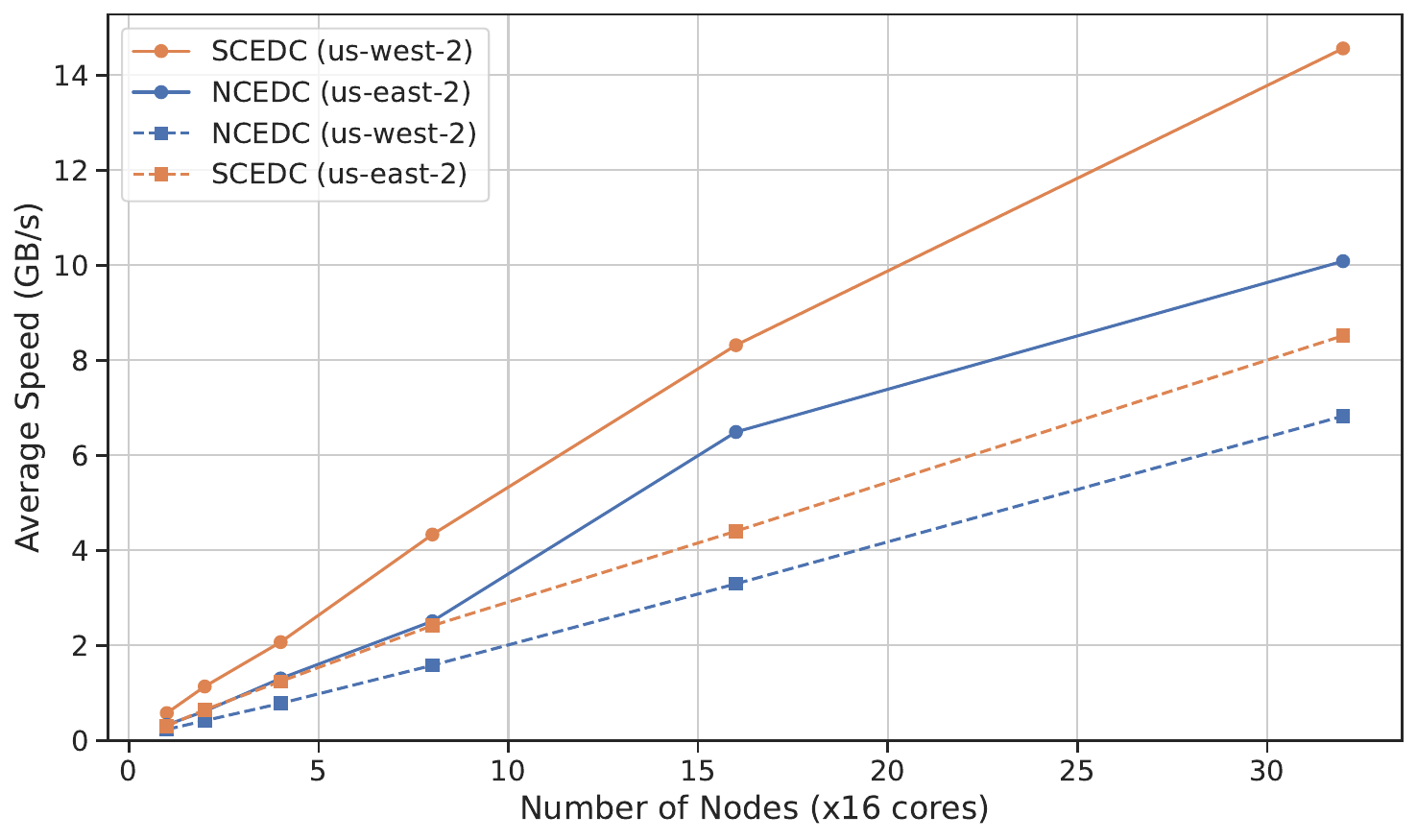}
    \caption{Average read speeds for accessing NCEDC and SCEDC AWS buckets within the same region (solid lines) and across different regions (dashed lines). The tests were conduced using the ``fsspec'' package on multiple 16-core AWS EC2 instances with results averaged over two repeated runs. The NCEDC and SCEDC archives are hosted in the us-west-2 and us-east-2 regions, respectively.}
    \label{fig:download_speed}
\end{figure}

An important application of deep learning in seismology involves mining seismic archives to detect hidden small earthquakes that conventional algorithms often miss.
Current seismic data mining faces significant challenges in downloading speed and storage space requirements for terabytes of continuous waveforms, along with the substantial computing resources needed for data processing.
Cloud computing provides an effective solution to both data access and processing challenges.
Continuous waveforms from Northern California and Southern California data centers are publicly hosted on AWS at \url{https://ncedc.org/db/cloud.html} and \url{https://scedc.caltech.edu/data/cloud.html}, comprising over 300 TB of data as of 2024 (\Cref{fig:aws_bucket}).
These cloud-hosted datasets serve as an excellent resource for large-scale seismic data analysis using cloud computing.
We demonstrate two methods for accessing AWS-hosted seismic datasets: direct mounting of AWS buckets and utilizing the unified file interface provided by \textit{fsspec}\footnote{\url{https://filesystem-spec.readthedocs.io/}}, as shown in the supplementary examples. Both methods are compatible with on-premises systems and cloud-based platforms.
Cloud computing provides various approaches for seismic data processing \citep{maccarthy2020seismology}, including virtual machines and containers for flexible computing capacity, serverless services like AWS Lambda for event-driven processing \citep{yu2021southern}, and batch processing services for large-scale parallel tasks typical in machine learning workflows \citep{krauss2023seismology}. Additionally, customized Kubernetes workflows can be deployed to orchestrate containerized applications, enabling efficient resource management, scalability, and portability \citep{zhu2023quakeflow}.
\Cref{fig:download_speed} shows the average reading speeds of SCEDC and NCEDC AWS buckets when accessed using \textit{fsspec} in a multi-node parallel configuration. The test data comprise 16,384 miniseed files from July 6–9, 2019, with dataset sizes of 110 GB for SCEDC and 80 GB for NCEDC. Reading from virtual machines located in the same AWS region as the buckets demonstrates significantly higher speeds compared to cross-region access. The performance difference between SCEDC and NCEDC may be attributed to regional variations in internet connection speeds or differences in bucket configurations. The SCEDC AWS bucket's longer public availability may contribute to its enhanced performance through more frequent usage and additional autoscaling resources.
Cloud computing significantly enhances data access speed,
, provides flexible computing resources, and dynamically scale computational capabilities with the demands of large-scale data mining and machine learning analysis.
This makes cloud computing a promising technique in modern seismology, facilitating a wide range of seismic data processing such as earthquake detection, event cataloging,  source characterization.

\section{Discussion}

The California earthquake event dataset (CEED) serves as a foundational resource for advancing machine learning development and seismic data mining tasks. By integrating datasets from both Northern and Southern California, this unified resource consolidates previous dataset formats and provides a consistent, robust resource for diverse machine learning research needs.
The dataset is designed for continuous updates as new data becomes available, enabling ongoing improvement of both the dataset and related machine learning models.
The dataset holds significant potential for generating improved seismic products, including high-resolution earthquake catalogs, focal mechanism solutions, and ground motion prediction models. These products, derived through continuously advancing deep learning models, could transform traditional cataloging approaches and provide new insights into earthquake processes.
The cloud-hosted continuous data archives further support data-intensive applications, such as searching for hidden earthquakes, conducting ambient noise analysis, and monitoring continuous velocity changes. 
These seismic data processing tasks are inherently parallelizable, making them well-suited for cloud computing to significantly enhance efficiency, scalability, and accessibility for large-scale seismic analysis.

While the growing volume of data benefits model training and validation capabilities, it also necessitates robust quality control mechanisms to maintain data integrity \citep{michelini2021instance}. 
The current dataset inevitably contains problematic labels due to human errors, noisy and ambiguous waveforms, inconsistent labeling standards, and missing labels for undetected events. 
Addressing these issues could enhance the reliability of models trained using the dataset. Future improvements would incorporate automated label correction mechanisms
to improve overall dataset quality and prevent bias in model training and application.
The current California earthquake dataset focuses primarily on seismometer waveforms, including broad-bands, strong motion sensors, and geophones. Expanding to include additional datasets, such as Distributed Acoustic Sensing (DAS) and GPS data, offers promising opportunities to broaden the dataset's applications. For example, DAS data can significantly enhance the spatiotemporal resolution in earthquake monitoring and fault zone structure studies \citep{zhan2020distributed,lindsey2021fiber}.
We have included a limited set of public DAS data from SCEDC \citep{yin2023earthquake}, formatted for training machine learning models such as PhaseNet-DAS \citep{zhu2023seismic} and available at \url{https://huggingface.co/datasets/AI4EPS/quakeflow_das}. 
Future efforts to incorporate new public DAS datasets such as the SeaFOAM DAS project \citep{romanowicz2023seafoam} will expand the dataset's utility beyond traditional seismic waveforms, making it a more comprehensive resource for multi-modal earthquake science research.

\section{Conclusions}

The California earthquake event dataset (CEED) provides a valuable resource for the continuous development of machine learning models and application of cloud computing to earthquake monitoring and seismic research. 
By integrating datasets from both Northern and Southern California, CEED leverages the long history and high quality of California's earthquake catalogs to provide a robust foundation for developing advanced deep learning models and driving progress toward next-generation artificial intelligence techniques for earthquake detection, characterization, and forecasting. 
The dataset also serves as a benchmark for evaluating deep learning model performance, improving the accuracy and reliability of seismic detection and interpretation.
Its open-access format and cloud computing compatibility facilitate continuous updates, reproducibility, and large-scale seismic data mining and analysis.
Integrated with other regional and global datasets, CEED would help advance comprehensive analysis of seismic activity and faulting processing and contributes to  seismic rick assessment along the San Andreas Fault System and other major fault systems globally.
Alongside other regional and global datasets, CEED contributes to a more comprehensive analysis of seismic activity and fault processes in California and worldwide.

\section{Acknowledgments}

We thank Gregory Beroza and Zachary Ross for their exploratory work in collecting datasets from Northern and Southern California, which contributed to the development of CEED.
The seismic data used in this study were collected by (1) the Berkeley Digital Seismic Network (BDSN, doi:10.7932/BDSN) and the USGS Northern California Seismic Network (NCSN, doi:10.7914/SN/NC) operated by \citet{BDSN} and \citet{NCSN}; and (2) the Southern California Seismic Network (SCSN, doi:10.7914/SN/CI) operated by \citet{SCSN}.
The original waveform data, metadata, and data products for this study were accessed through the Northern California Earthquake Data Center (doi:10.7932/NCEDC) \citep{ncedc2014northern} and the Southern California Earthquake Center (doi:10.7909/C3WD3xH1) \citep{scedc2013southern}.
Please include acknowledgment and citation of the original data providers when using this dataset.
This research was supported by the Statewide California Earthquake Center (Contribution No.24133). SCEC is funded by NSF Cooperative Agreement EAR-2225216 and USGS Cooperative Agreement G24AC00072-00.

\clearpage
\bibliography{references}

\clearpage
\appendix


\section*{Supplementary material}

# California *E*arthquake *E*vent *D*ataset for Machine Learning and Cloud Computing (CEED)

- Load necessary libraries and ultility functions

```
In [1]: import os
        import obspy
        import fsspec
        import pandas as pd
        from glob import glob
        import time
        import matplotlib.pyplot as plt
        import datasets
        import numpy as np

        result_path = "results"
        figure_path = "figures"
        if not os.path.exists(result_path):
            os.makedirs(result_path)
        if not os.path.exists(figure_path):
            os.makedirs(figure_path)


        def map_cloud_path(root_path, provider, starttime, network, station, location, channel):
            if isinstance(starttime, str):
                starttime = pd.Timestamp(starttime)
            if provider.lower() == "scedc":
                year = starttime.strftime("%Y")
                dayofyear = starttime.strftime("%j")
                if location == "":
                    location = "__"
                path = f"{root_path}/{provider.lower()}-pds/continuous_waveforms/{year}/{year}_{dayofyear}/{network}{station:_<5}{channel}{location:_<2}_{year}{dayofyear}.ms"
            elif provider.lower() == "ncedc":
                year = starttime.strftime("%Y")
                dayofyear = starttime.strftime("%j")
                path = f"{root_path}/{provider.lower()}-pds/continuous_waveforms/{network}/{year}/{year}.{dayofyear}/{station}.{network}.{channel}.{location}.D.{year}.{dayofyear}"
            else:
                raise ValueError(f"Unknown provider: {provider}")
            return path
```

## On-prem access to the NCEDC and SCEDC s3 buckets

```
In [2]: mseed_list = [
            {
                "provider": "scedc",
                "network": "CI",
                "station": "CCC",
                "location": "",
                "channel": "HHZ",
                "year": "2019",
                "month": "07",
                "day": "06",
            },
            {
                "provider": "ncedc",
                "network": "BK",
                "station": "CMB",
                "location": "00",
                "channel": "HHZ",
                "year": "2019",
```

```
        "month": "07",
        "day": "06",
    },
]
```

### Mount the NCEDC and SCEDC s3 buckets as local directories

```
mkdir -p $HOME/cloud/scedc-pds
mkdir -p $HOME/cloud/ncedc-pds

s3fs scedc-pds $HOME/cloud/scedc-pds -o allow_other -o public_bucket=1 -o compat_dir
s3fs ncedc-pds $HOME/cloud/ncedc-pds -o allow_other -o public_bucket=1 -o compat_dir
```

In [3]:
```python
root_path = os.path.expanduser("~/cloud")
for mseed_info in mseed_list:
    starttime = pd.Timestamp(f"{mseed_info['year']}-{mseed_info['month']}-{mseed_info['day']}T00:00:00")
    file_path = map_cloud_path(
        root_path,
        mseed_info["provider"],
        starttime,
        mseed_info["network"],
        mseed_info["station"],
        mseed_info["location"],
        mseed_info["channel"],
    )
    t0 = time.time()
    stream = obspy.read(file_path)
    print(f"Reading {file_path}: {time.time() - t0:.1f}s")
    stream.plot(outfile=f"{figure_path}/{file_path.split('/')[-1]}.png")
```

```
Reading /Users/weiqiang/cloud/scedc-pds/continuous_waveforms/2019/2019_187/CICCC__HHZ___2019187.ms: 3.4s
Reading /Users/weiqiang/cloud/ncedc-pds/continuous_waveforms/BK/2019/2019.187/CMB.BK.HHZ.00.D.2019.187: 6.2s
```

### Direct access to the s3 buckets using fsspec

In [4]:
```python
root_path = "s3:/"
for mseed_info in mseed_list:
    file_path = map_cloud_path(
        root_path,
        mseed_info["provider"],
        starttime,
        mseed_info["network"],
        mseed_info["station"],
        mseed_info["location"],
        mseed_info["channel"],
    )
    t0 = time.time()
    with fsspec.open(file_path, s3={"anon": True}) as f:
        stream = obspy.read(f)
    print(f"Reading {file_path}: {time.time() - t0:.1f}s")
    stream.plot(outfile=f"{figure_path}/{file_path.split('/')[-1]}.png")
```

```
Reading s3://scedc-pds/continuous_waveforms/2019/2019_187/CICCC__HHZ___2019187.ms: 1.6s
Reading s3://ncedc-pds/continuous_waveforms/BK/2019/2019.187/CMB.BK.HHZ.00.D.2019.187: 2.2s
```

## Machine Learning

### Example: Reading the test datasets from Hugging Face.

By default, we select the most recent year of the dataset as the test dataset and the preceding year as the training dataset.

The training dataset is large exceeding 800GB, here we will use the test dataset for this example.

```python
In [12]: # %%
         quakeflow_nc = datasets.load_dataset("AI4EPS/CEED", name="station_test", split="test")

         for example in quakeflow_nc:
             print(example.keys())
             for key in example.keys():
                 if key == "data":
                     print(key, np.array(example[key]).shape)
                 else:
                     print(key, example[key])
             break
```

```
Loading dataset shards:   0%|          | 0/56 [00:00<?, ?it/s]
dict_keys(['data', 'phase_time', 'phase_index', 'phase_type', 'phase_polarity', 'begin_time', 'end_t
ime', 'event_time', 'event_time_index', 'event_location', 'station_location'])
data (3, 8192)
phase_time ['2023-01-01T06:49:13.520000', '2023-01-01T06:49:15.270000', '2023-01-01T06:49:19.14000
0']
phase_index [3335, 3510, 3897]
phase_type ['P', 'P', 'S']
phase_polarity ['U', 'U', 'N']
begin_time 2023-01-01T06:48:40.170000
end_time 2023-01-01T06:50:40.170000
event_time 2023-01-01T06:49:08.370000
event_time_index 2820
event_location [-121.19933319091797, 36.595333099365234, 8.399999618530273]
station_location [-121.44721984863281, 36.76403045654297, -0.3172000050544739]
```

## Example: Phase Picking using PhaseNet

```python
In [7]: ## Download code from github
        !git clone git@github.com:AI4EPS/EQNet.git
```

```
fatal: destination path 'EQNet' already exists and is not an empty directory.
```

```python
In [8]: ## Prepare a list of mseed files
        mseed_list = [
            [
                {
                    "provider": "scedc",
                    "network": "CI",
                    "station": "CCC",
                    "location": "",
                    "channel": ch,
                    "year": "2019",
                    "month": "07",
                    "day": "06",
                }
                for ch in ["HHZ", "HHE", "HHN"]
            ],
            [
                {
                    "provider": "ncedc",
                    "network": "BK",
                    "station": "CMB",
                    "location": "00",
                    "channel": ch,
                    "year": "2019",
                    "month": "07",
                    "day": "06",
                }
                for ch in ["HHZ", "HHE", "HHN"]
            ],
        ]

        root_path = "s3:/"   # cloud
        # root_path = os.path.expanduser("~/cloud") # local
        with open("mseed.txt", "w") as fp:
            for station in mseed_list:
```

```python
            file_paths = []
            for i, mseed_info in enumerate(station):
                starttime = pd.Timestamp(f"{mseed_info['year']}-{mseed_info['month']}-{mseed_info['day']}T00:00:00")
                file_path = map_cloud_path(
                    root_path,
                    mseed_info["provider"],
                    starttime,
                    mseed_info["network"],
                    mseed_info["station"],
                    mseed_info["location"],
                    mseed_info["channel"],
                )
                file_paths.append(file_path)
            fp.write(",".join(file_paths) + "\n")
```

In [9]:
```
## Check the mseed list file
!head mseed.txt
```

s3://scedc-pds/continuous_waveforms/2019/2019_187/CICCC__HHZ___2019187.ms,s3://scedc-pds/continuous_waveforms/2019/2019_187/CICCC__HHE___2019187.ms,s3://scedc-pds/continuous_waveforms/2019/2019_187/CICCC__HHN___2019187.ms
s3://ncedc-pds/continuous_waveforms/BK/2019/2019.187/CMB.BK.HHZ.00.D.2019.187,s3://ncedc-pds/continuous_waveforms/BK/2019/2019.187/CMB.BK.HHE.00.D.2019.187,s3://ncedc-pds/continuous_waveforms/BK/2019/2019.187/CMB.BK.HHN.00.D.2019.187

In [10]:
```python
## Run PhaseNet on the mseed files
cmd = f"python EQNet/predict.py --model phasenet --data_list mseed.txt --result_path {result_path} --batch_size=1 --format mseed --device cpu"
os.system(cmd)
```

Not using distributed mode
Namespace(model='phasenet', resume='', backbone='unet', phases=['P', 'S'], device='cpu', workers=0, batch_size=1, use_deterministic_algorithms=False, amp=False, world_size=1, dist_url='env://', data_path='./', data_list='mseed.txt', hdf5_file=None, prefix='', format='mseed', dataset='das', result_path='results', plot_figure=False, min_prob=0.3, add_polarity=False, add_event=False, sampling_rate=100.0, highpass_filter=0.0, response_path=None, response_xml=None, subdir_level=0, cut_patch=False, nt=20480, nx=5120, resample_time=False, resample_space=False, system=None, location=None, skip_existing=False, distributed=False)
Total samples:  ./mseed : 2 files
Predicting: 100%|██████████| 2/2 [00:58<00:00, 29.20s/it]
Merging picks_phasenet: 2it [00:00, 128.78it/s]
Number of picks: 9471
Number of P picks: 5306
Number of S picks: 4165

Out[10]: 0

In [11]:
```python
## Check out prediction results
if not os.path.exists(figure_path):
    os.makedirs(figure_path)
for station in mseed_list:
    mseed_info = station[0]
    starttime = pd.Timestamp(f"{mseed_info['year']}-{mseed_info['month']}-{mseed_info['day']}T00:00:00")
    file_path = map_cloud_path(
        root_path,
        mseed_info["provider"],
        starttime,
        mseed_info["network"],
        mseed_info["station"],
        mseed_info["location"],
        mseed_info["channel"],
    )
    picks = pd.read_csv(f"{result_path}/picks_phasenet/{file_path.split('/')[-1]}.csv", parse_dates=["phase_time"])

    plt.figure()
    plt.hist(picks[picks["phase_type"] == "P"]["phase_time"], bins=100, alpha=0.5, label="P")
    plt.hist(picks[picks["phase_type"] == "S"]["phase_time"], bins=100, alpha=0.5, label="S")
```

```
    plt.legend()
    plt.xlabel("Time")
    plt.ylabel("Frequency")
    plt.title(f"{file_path.split('/')[-1]}")
    plt.gcf().autofmt_xdate()
    plt.savefig(f"{figure_path}/{file_path.split('/')[-1]}_picks.png")
```

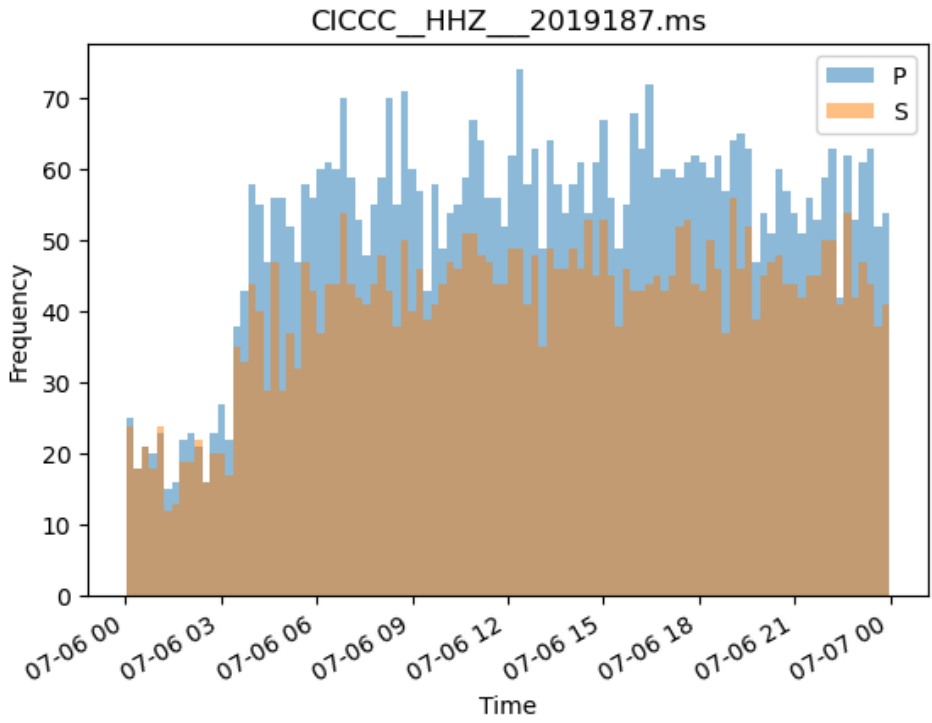

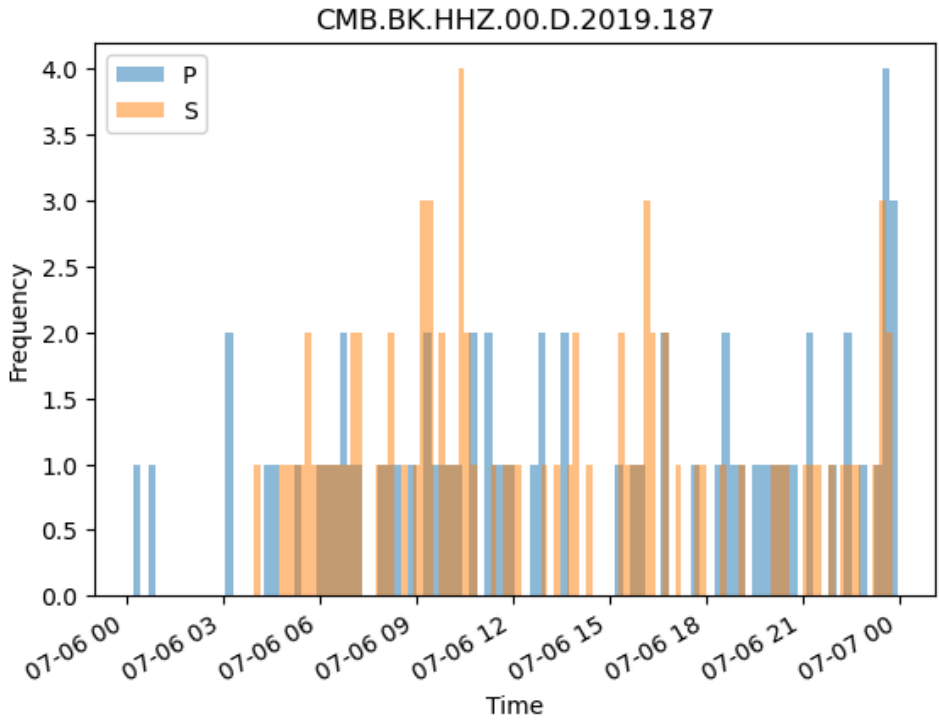



\end{document}